\documentclass[nologo,11pt,a4paper]{ETHpaper}
\usepackage{graphicx, amsmath, amssymb,color,wasysym}
\usepackage{subcaption}

\usepackage{gensymb} \usepackage{appendix}

\usepackage[square,numbers,sort&compress]{natbib}

\usepackage{tikz}
\usepackage{tikz-3dplot}

\begin{document}

\newcommand{\mean}[1]{\left\langle #1 \right\rangle} 
\newcommand{\abs}[1]{\left| #1 \right|} 
\newcommand{\ul}[1]{\underline{#1}}
\renewcommand{\epsilon}{\varepsilon} 
\newcommand{\eps}{\varepsilon} 
\renewcommand*{\=}{{\kern0.1em=\kern0.1em}}
\renewcommand*{\-}{{\kern0.1em-\kern0.1em}} 
\newcommand*{\+}{{\kern0.1em+\kern0.1em}}

\newcommand{\RA}{\Rightarrow}
\newcommand{\bbox}[1]{\mbox{\boldmath $#1$}}

\title{An agent-based model of multi-dimensional opinion dynamics and opinion alignment}

\titlealternative{An agent-based model of multi-dimensional opinion dynamics and opinion alignment}

\author{Simon Schweighofer$^{1}$, David Garcia$^{1}$, Frank Schweitzer$^{1,2\star}$}
\footnotetext{$^{\star}$ Corresponding author: \texttt{fschweitzer@ethz.ch}}
\authoralternative{S. Schweighofer, D. Garcia, F. Schweitzer}

\address{$^{1}$ Complexity Science Hub Vienna \\
  $^{2}$Chair of Systems Design, ETH Zurich, Weinbergstrasse 58, 8092 Zurich, Switzerland}

\reference{(Submitted for publication)} 
\www{\url{http://www.sg.ethz.ch}}

\makeframing
\maketitle

\begin{abstract}
  It is known that individual opinions on different policy issues often align to a dominant ideological dimension (e.g. ``left'' vs. ``right'') and become increasingly polarized.
  We provide an agent-based model that reproduces these two stylized facts as emergent properties of an opinion dynamics in a  multi-dimensional space of continuous opinions.
  The mechanisms for the change of agents' opinions in this multi-dimensional space are derived from cognitive dissonance theory and structural balance theory.
  We test assumptions from proximity voting and from directional voting regarding their ability to reproduce the expected emerging properties.
  We further study how the emotional involvement of agents, i.e. their individual resistance to change opinions, impacts the dynamics.
  We identify two regimes for the global and the individual alignment of opinions.
  If the affective involvement is high and shows a large variance across agents, this fosters the emergence of a dominant ideological dimension.   
  Agents align their opinions along this dimension in opposite directions, i.e. create a state of polarization.
  
  \end{abstract}

\section{Introduction}
\label{sec:introduction}

The famous economist Nicolas Kaldor in 1961 suggested that  theorists ``should be free to start off with a stylised view of the facts – i.e. concentrate on broad tendencies, ignoring individual detail'' \citep{kaldor1961capital}.
His advice was certainly taken by the numerous physicists modeling opinion dynamics \citep{castellano2009statistical,holyst2001social,Galam_2000}, one of the most flourishing topics in the area of \emph{sociophysics} \citep{Schweitzer2018}.
In many of theses models, opinions are treated as binary variables, $\{0,1\}$, very much like spins, and changes in opinions follow rather simplistic rules.
Despite their abstract nature, these models have generated interesting insights into the dynamics of disordered systems \citep{dornic2001critical,redner-03-a,min2017fragmentation,tessone2008a}.
For example, voter models allowed to study under which conditions ``consensus'', i.e. a large domain with aligned spins, can be obtained, or how a ``minority'' and a ``majority'' can coexist \citep{artime2019herding,carro2016noisy,fs-voter-03}. 

The question is how well such models fare with respect to \emph{real} opinion dynamics \citep{fernandez2014voter}.
To answer it in the spirit of Kaldor requires us to specify the ``stylized facts'' that shall be used as a ground truth, or a reference for judging such models.
While there is no common agreement on these stylized facts, we can certainly pick, from our everyday experience, two observations in political space  that most scholars would subscribe to:
(i) Opinions have become increasingly \emph{polarized}, i.e. there are two fractions of almost equal share in the population with opposite opinions \citep{garcia2019,schweighofer2019,wilford2017polarization,grechyna2016determinants,bornschier2015new,iyengar2015fear,garcia2015ideological,suhay2015explaining,dimock2014political,druckman2013elite,mason2013rise,garcia2012b,conover2012partisan,hetherington2009review}.
(ii) This polarization holds regardless of a specific topic, but is attributed to rather general \emph{ideological positions}, for example ``democrats'' vs. ``republicants'' in the U.S., or ``conservative'' vs. ``liberal'', or ``left'' vs. ``right'' in Europe \citep{garcia2015ideological,mason2015distinguishing,kimfordingpowell2010,leimgruber2010comparing,malka2010more}.

Can the voter-type opinion dynamic models proposed by socio-physicists replicate such stylized facts?
Yes, but in a rather trivial manner.
Polarization is built into the models by the dichotomy of the two opposite opinions.
So, if we do not obtain consensus in the long run, i.e. the dominance of one opinion, we obtain polarization, i.e. the coexistence of two ``extreme'' opinions.
In the absence of any alternative, these opinions already represent the ideological ``left'' and ``right'' positions. 
Model parameters allow us to adjust the fractions of the respective camps even to 50/50, i.e. a stalemate reminiscent of real political situations in quite a number of different countries.

We argue that such models have \emph{not} passed the test, for obvious reasons:
They do not show the \emph{emergence} of a polarized opinion state and they also do not show the \emph{emergence} of a ideological dimension along which polarization occurs.
The term \emph{emergence} refers to a process of self-organization that leads to a new \emph{systemic} property as the result of the dynamic interactions between a large number of individuals.
In our paper these individuals are  represented as agents with certain internal degrees of freedom, most notably their opinion.
Interactions refer to the exchange of information about the opinions of others, which in turn results in an adjustment of the opinion of each individual.
With respect to opinion dynamics, the emerging property is the ideological dimension.
Agents should \emph{align} their opinions on this ideological dimension.
Polarization then means that they align their opinions in \emph{opposite directions}, whereas \emph{consensus} means that the align their opinion in the \emph{same direction}.

In order to obtain such emerging properties, we have to change  from \emph{binary} opinions to \emph{continuous} opinions which follow some distribution.
These opinions can still be mapped to a finite interval, e.g. $[-1,+1]$, but extreme opinions should be less frequent, at least initially, than ``normal'' ones.
Secondly, we have to change from \emph{one-dimensional} opinions to \emph{multi-dimensional} opinions.
Each dimension represents for example a given policy issue about which an individual can have its own opinion. 
The dominating ideological dimension is not one of these named policy dimensions but a combination of all of these.
It shall be the one that classifies these multi-dimensional opinions best, on a one-dimensional scale.
``Best'' means to explain most of the variance of all multi-dimensional opinions of a large number of individuals.
Our paper will address specifically this last point in Section~\ref{sec:alignm-princ-comp}.

But there is more to it.
Experienced modelers would probably know how to obtain the requested outcome from simplistic assumptions.
However, even a ``correct'' outcome on the macro level does \emph{not} allow us to conclude that the respective assumptions for interactions on the micro level are ``correct'' as well.
Because there are various ways of obtaining a given outcome, we need additional evidence for our micro mechanisms.
That means, we have to base our  interaction model on  \emph{theories} or \emph{experiments} that justify our assumptions.
This is the most neglected problem of socio-physics models of opinion dynamics.
To solve it would require to learn about works in sociology, psychology, political science, to consider \emph{why} and \emph{how} individuals change their opinions.
These insights can still be formalized as shown for example in \citep{schweitzer2014c}.
The rules for interactions are then no longer ad-hoc assumptions, but backed up by additional disciplinary arguments.

Our paper will address this problem in the following Section~\ref{sec:probl-multi-dimens}.
We then continue with introducing our agent-based model in multi-dimensional opinion space of $m=\{1,2,...,M\}$ policy issues  in Section~\ref{sec:multi-dimens-opin}.
For the opinion dynamics, we implement two different paradigms, \emph{proximity voting} and \emph{directional voting} and explain their
shortcomings in Sections~\ref{sec:multi-dimens-opin}, \ref{sec:alig}.
Based on these insights, in Section~\ref{sec:abm_3} we provide a model of multi-dimensional opinion dynamics that is able to produce the desired outcome mentioned above in a robust manner.
This model also considers the emotional involvement of agents when changing their opinions.
Therefore, in Section~\ref{sec:alignm-princ-comp} we investigate in detail how the emergence of a dominating ideological dimension depends on this emotional involvement.
Section~\ref{sec:abm_conclusion} concludes with general reflections on agent-based modeling and on the achievements obtained from our multi-dimensional modeling approach.

\section{The problems of multi-dimensional opinion dynamics}
\label{sec:probl-multi-dimens}

\subsection{Combining Cognitive Dissonance and Structural Balance}
\label{sec:comb-cogn-diss}

\paragraph{Micro mechanisms. \ }

Our first aim is to motivate our rules of opinion change from a plausible set of micro-mechanisms \citep{hedstrom2005dissecting}. These micro mechanisms are derived from established psychological theories, in particular \emph{cognitive dissonance theory} \citep{heider1946attitudes,festinger1962theory}, and its extension to social relations, \emph{structural balance theory} \citep{cartwright1956structural}.

Cognitive dissonance theory focuses on the perspective of the \emph{individual}, specifically the relation between its beliefs, or opinions.
It postulates that, if an individual holds two or more beliefs that it judges as contradictory, it will experience this as unpleasant. 
To alleviate this unpleasantness, the individual will either adapt or drop one of these beliefs to re-establish accordance.
Hence, an individual has the tendency to minimize cognitive dissonance, which can be seen as a micro-foundation of opinion formation \citep{schweitzer2014c}.

\paragraph{Stable and unstable triads. \ }

Structural balance theory extends cognitive dissonance theory to explain the \emph{relations between individuals}.
Two individuals $i$ and $j$ can have either a positive relation, $r^{ij}=+1$ or a negative one, $r^{ij}=-1$.
Structural balance theory focuses on \emph{triadic} relations $\{i,j,k\}$, i.e. relations between three individuals $i$, $j$, and $k$. 
It postulates that there are stable and unstable triads.
Whether a triad is stable or unstable can be determined by multiplying the signs of the relations $r^{ij}$, $r^{ik}$ and $r^{jk}$ in the triad.
For example, if individuals $i$, $j$, and $k$ have exclusively positive relations with each other, $r^{ij}=r^{ik}=r^{jk}=+1$, the product of their signs would be positive, i.e. the triad is stable.
If, however, two relations are positive and one is negative, e.g.  $r^{ij}=r^{ik}=+1$, $r^{jk}=-1$, the product would be negative and the triad is assumed to be unstable.
Unstable triads have the tendency to transform themselves into stable triads.
This means that either $i$ convinces $j$ and $k$ to change their relations into a positive one, $r^{jk}=+1$, or $i$ changes its own relation to either $j$ or $k$ to a negative one, $r^{ij}=-1$ \emph{or} $r^{ik}=-1$.
In the latter case, the triad would have two negative relations and one positive.
Hence, the product of the signs becomes positive and the triad has become stable.

\paragraph{Extension to opinion dynamics. \ }

We now combine the assumptions of both theories, to explain how two individuals adjust their opinions.  
We build on our previous work on \textit{weighted balance theory} \cite{schweighofer2019}, which combines elements  of cognitive balance theory with an emotional factor of evaluative extremeness that was calibrated through the empirical analysis of an electoral survey. 

\begin{figure}[htbp]
        \includegraphics[width=0.4\textwidth]{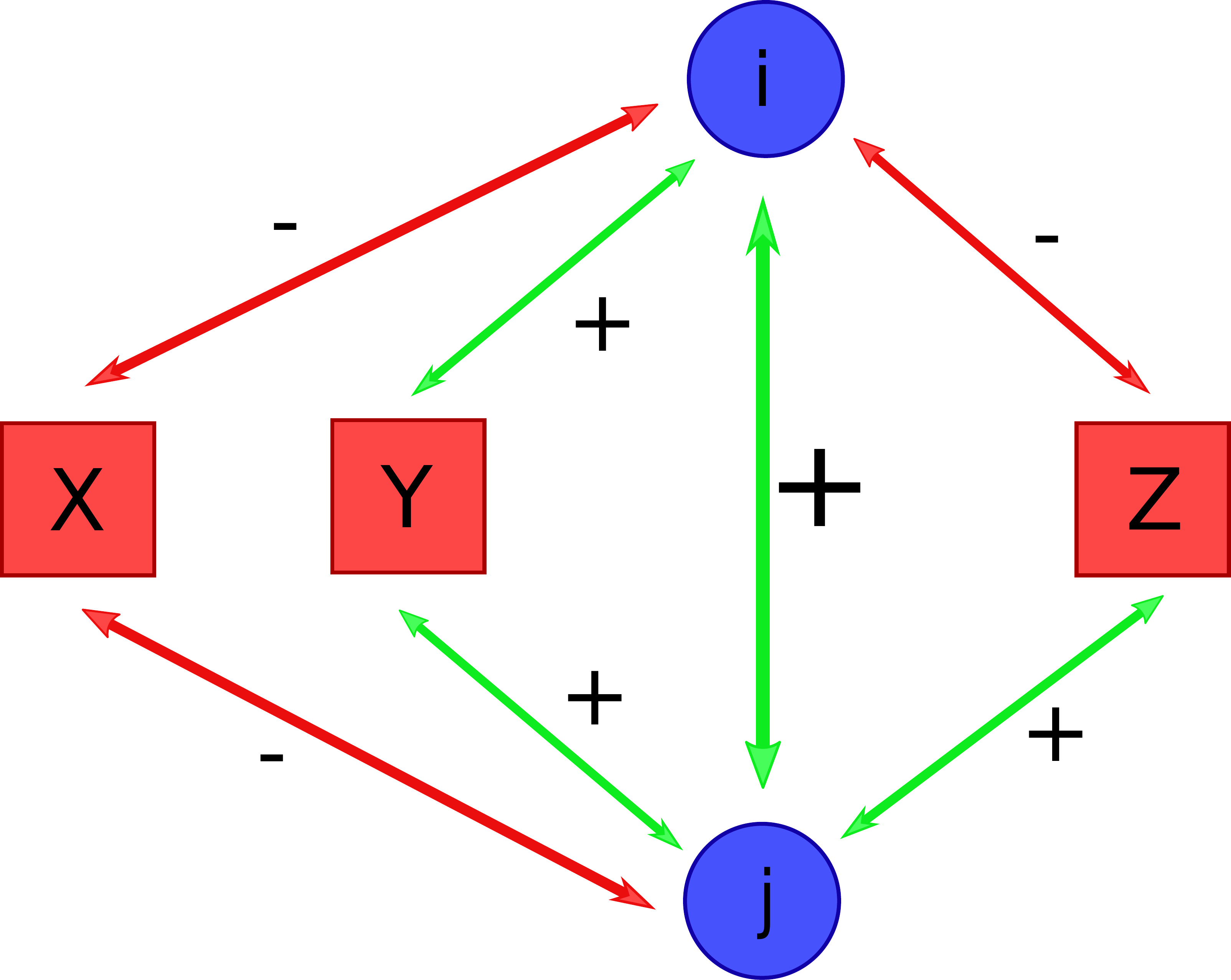}(a)
\hfill
        \includegraphics[width=0.4\textwidth]{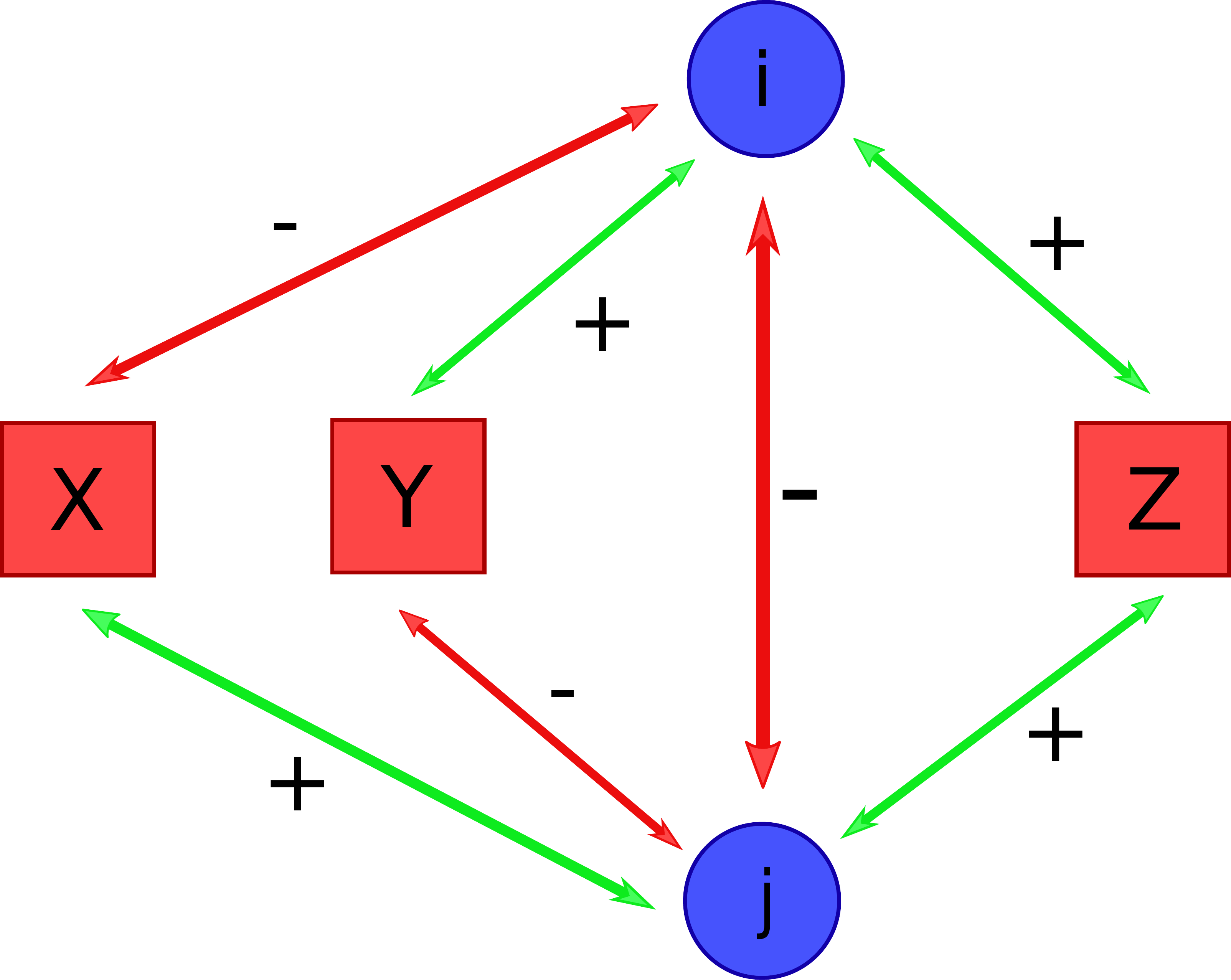}(b)
    \caption{Two agents $i$ and $j$ agreeing on policy dimension $x$ and $y$, but disagreeing on $z$ (left panel), and disagreeing on $x$ and $y$, but agreeing on $z$ (right panel).}
    \label{fig:struct_bal}
\end{figure}

Figure \ref{fig:struct_bal} considers two individuals $i$ and $j$ and their opinions on three policy issues $x$, $y$, and $z$.
For the moment we assume that each individual can have only have a positive or negative stance on each issue.
Their opinions can be expressed by the opinion vectors $\mathbf{o}^{i}=\{o^{i}_{x},o^{i}_{y},o^{i}_{z}\}$. 
In the example shown in Figure \ref{fig:struct_bal}(a), for $i$ we find that $o^{i}_{x}=o^{i}_{z}=-1$, $o^{i}_{y}=+ 1$, while for $j$ we find that
$o^{j}_{x}=-1$, $o^{j}_{y}=o^{j}_{z}=+1$.
That means, both individuals have a negative stance on issue $x$ and a positive stance on $y$, but on issue $z$ their opinions contradict each other.

Regarding the \emph{relation between individuals $i$ and $j$}, we now make the following assumption: 
Since $i$ and $j$ agree on two issues and disagree only on one, they have a positive relation, i.e. $r^{ij}=+1$.
Considering this relation between $i$ and $j$, now three different triads result, each of which contains $i$, $j$, and one of the three policy issues, $x$, $y$ or $z$.
Based on the rules explained above, we see that the triads $\{i,j,x\}$ and $\{i,j,y\}$ are \emph{stable} because the product of the signs is positive, whereas  the triad $\{i,j,z\}$ is \emph{unstable} because the product of the signs is negative.

This instable triad will induce a cognitive dissonance in both $i$ and $j$. 
Individual $i$ might ask herself "How can it be that I agree with $j$ on most issues, but disagree on $z$?"
To resolve the dissonance, $i$ will either adapt her own opinion towards $j$, $o^{i}_{z}\to o^{j}_{z}$, or she will try to convince $j$ to adapt her opinion, $o^{i}_{z}\gets o^{j}_{z}$. 
Either way, $i$ and $j$ will strive to come to an \emph{agreement} with respect to their opinions on issue $z$.
We can see this as an \emph{attractive force} between the opinion vectors of $i$ and $j$, which eventually leads to a further alignment of opinions.
We will make use of this consideration in Sections \ref{sec:alig} and \ref{sec:abm_3}.

\subsection{Constraints on interactions}
\label{sec:restr-inter}

The above example assumes that the two individuals agree on most issues and therefore have an incentive to resolve discrepancies for those cases where they don't agree.
But what happens if individuals already disagree on most issues and only agree on very few?

In the simplest case, which is also used in the bounded confidence model discussed in Section \ref{sec:simpl-probl}, one can assume that these individuals do \emph{not interact} anymore if differences in their opinions are larger than a certain \emph{threshold} $\epsilon$. 
This seems to be a reasonable argument because, without interaction, they are no longer confronted with the cognitive dissonance resulting from it. 
This argument also underlies the much-discussed \emph{filter bubble} \citep{pariser2011filter}, which emerges if filter algorithms in online media shield users from information and opinions which do not fit their own.
While empirical evidence shows that social media encourages more diverse news diets \cite{Scharkow2020}, selective media exposure \cite{Bakshy2015} can serve as an interaction boundary that contributes to polarization of opinions by preventing interaction across long ideological distances.

We will investigate the impact of interaction thresholds for the opinion dynamics in Sections \ref{sec:multi-dimens-opin} and \ref{sec:alig}, but already here we want to point to some problems involved in this assumption.
To define a critical distance between opinions is appropriate if one considers a one-dimensional opinion space, as most opinion dynamics models do \citep{lorenz2007continuous}. 
But how should we interpret this distance in a multi-dimensional opinion space, where the opinions of two individuals regarding an issue $x$ are very close, but regarding an issue $z$ are very different?
Will these individuals interact or not? 
Averaging over different opinion dimensions would make no sense.
This problem will be further discussed in Sections \ref{sec:simpl-probl} and \ref{sec:subj-simil-opin}. 

Secondly, the interaction threshold $\epsilon$ is usually assumed to be constant and equal across individuals.
The bounded confidence model, for example, simply treats this as a tunable parameter that impacts the possibility of reaching consensus.
But it is very important for the opinion  dynamics  to what extent an individual may be affected by the respective policy issues. 
Psychological research shows that opinions with a stronger emotional component are more \emph{resistant} to change \citep{schank2013beliefs}.
In other words, beliefs or opinions associated with strong emotional reactions are more stable.
Hence, this individual emotional level may have an impact on the interaction threshold, i.e. whether or nor opinions change.  
The more an individual is emotionally involved in, e.g., politics, the sooner she will experience the critical level of cognitive dissonance that makes her abort or avoid interaction.
Therefore, in Sections \ref{sec:alig} and \ref{sec:abm_3} we will discuss how the emotional involvement of an individual can be related to her interaction threshold.

\subsection{Increasing disalignment}
\label{sec:incr-disal}

If individuals already disagree on most issues and only agree on very few, they also have the possibility to \emph{adjust} their opinions, but into the \emph{negative} direction.
That means, they resolve their cognitive dissonance by also disagreeing on the few issues, they previously had agreed on.
In Figure \ref{fig:struct_bal}(b), we consider the example that individuals 
$i$ and $j$ disagree on issues $x$ and $y$, and only agree on $z$.
Because they disagree on more issues than they agree, the overall relation between $i$ and $j$ in this case is negative, $r^{ij}=-1$.
Considering this relation between $i$ and $j$, the  triads $\{i,j,x\}$ and $\{i,j,y\}$ are \emph{stable}, because $i$ and $j$ disagree on the issues $x$ and $y$.
However, the triad $\{i,j,z\}$ is \emph{unstable} because   
$i$ and $j$, even though they have a negative relation with each other, agree on issue $z$. 
This triad produces a cognitive dissonance for $i$ and $j$, which has to be resolved in some way.

Instead of simply stopping their interaction, individual $i$ can also change her opinion $o^{i}_{z}=+1$ to $o^{i}_{z}=-1$.
This transforms the triad $\{i,j,z\}$ into a stable one, the same would result from $j$ changing her opinion $o^{j}_{z}$. 
As a result, $i$ and $j$ are now in disagreement on all three issues $x$, $y$ and $z$.
But because of the negative relation $r^{ij}=-1$, their cognitive dissonances are reduced to a minimum.
This outcome, which is desirable for both individuals, postulates a \emph{repulsive force} between the political positions of $i$ and $j$. 

This seems counter-intuitive only if we assume that, if two individuals interact, they should end up agreeing on more issues than before, and not on less. 
Instead, there is empirical evidence \cite{hovland1957assimilation} that, when individuals are confronted with positions on alcohol prohibition that they fundamentally disagree with, they move away from these positions. 
Similarly, in an experiment \cite{druckman2013elite} partisan voters were confronted with the information that leaders of the opposite party endorse a certain policy and then adjusted their political position so as to contradict this policy. 

The existence of negative social influence is still debated in the literature.
A recent study \cite{takacs2016discrepancy} could not find any evidence for negative influence, but only tested social influence on a single issue dimension. 
This might not be applicable to our scenario with several opinion dimensions.
Hence, in Section \ref{sec:abm_3} we test the modeling assumption that, if two agents disagree on too many issues, they modify their respective opinions such that they increase their overall disagreement.

\subsection{Dimensionality reduction of opinion space}
\label{sec:dimens-reduct-opin}

In this paper, we consider a multi-dimensional opinion space.
Each dimension represents a different issue, on which a given individual can have her own opinions.
However, we cannot assume that her opinions are completely independent of each other.
Instead, it is known that opinions on policy issues are correlated \citep{benoit2012dimensionality,benoit2006party}.
For example, individuals with a positive stance on cannabis legalization more likely have a negative stance on nuclear energy.  
Political scientists call the correlation among opinions \emph{issue constraint} (or issue alignment) \citep{converse1964nature, dimaggio1996have, baldassarri2008partisans} because 
it basically constraints the possible combinations of opinions on different issues.

That implies the number of independent dimensions to describe all opinions is effectively reduced, i.e. opinion dimensions are ``bundled'' \citep{poole2005spatial}.
From the individual perspective, it means that opinions on different issues are \emph{aligned}.
Remarkably, from the systemic perspective of all individuals, it was found that most individuals align their opinion in the same manner,  i.e. one observes the emergence of a dominant dimension that explains most of the variability of individual opinions on different issues.
In political science, this main ``ideological'' dimension is called the ``left-right'' dimension, in the US also the ``liberal-conservative'' dimension \citep{benoit2006party}.
To characterize the opinion spectrum of individuals regarding different issues, it is sufficient to position them on this \emph{continuous} ``left-right'' axis.
This does not imply that \emph{all} opinions are explained this way, but \emph{most} of their variance is.

Political scientists have applied this dimensionality reduction to survey data or voting records \citep{poole2005spatial,benoit2012dimensionality,benoit2006party}. 
Studies have shown that political behavior and decision making, such as election choices, legislative decisions, coalition formation, and judicial decisions, can be explained to a large degree based on the ideological positions of political actors \citep{merrill1999unified,Indridason2011}.

The emergence of this dominant dimension is denoted as \emph{global alignment} in political science and addressed as a major open research question \citep{poole2005spatial}.
It was argued that, in addition to extreme opinions, high global alignment is a necessary condition to observe political \emph{polarization} \citep{ dimaggio1996have, baldassarri2008partisans}.
In fact, without global alignment individuals with extreme opinions on some dimensions could still find themselves close to the opinions of others on other dimensions.
This would then allow them to still interact, despite their discrepancies.
But the emergence of global alignment enforces the characterization of individual opinions on just one ``left-right'' axis.
The opinion dynamics, with respect to this dominating dimension, then no longer enables individuals to interact, because their distance along this dimension is too large.

It is a major goal of our paper to provide an agent-based model of opinion dynamics that is able to reproduce the emergence of global alignment, starting from a random distribution of opinions.
While the existence of alignment is already discussed particularly in political science, the challenge is to present a model that can generate this global alignment as an emerging phenomenon, without encoding it in the setup of the model.
Hence, we will evaluate the outcome of our simulations with respect to this very challenge, in Section~\ref{sec:alignm-princ-comp}.

Additionally, we are interested in the relation between global alignment and individual affective involvement into politics, which is
is seen as a precondition for the emergence of global alignment.
Based on the fact that political elites usually exhibit much stronger issue alignment than the general population, \citet[p.211][]{poole2005spatial} believes that ``part of the answer to these questions is that political elites are \emph{passionate} about their beliefs''.
Therefore, in Section \ref{sec:alignm-princ-comp} we want to test whether, on the systemic level,  
a higher average affective involvement will produce higher \emph{global alignment} and, on the individual level, a higher individual involvement should exhibit higher \emph{individual alignment}.

\section{Modeling multi-dimensional opinion dynamics}
\label{sec:multi-dimens-opin}

\subsection{Agent-based model of opinion dynamics}
\label{sec:agent-based-model}

\paragraph{Agent variables. \ }

In the following we consider a \emph{multi-dimensional opinion space}, in which each dimension $m=\{1,2,...,M\}$ refers to a specific political issue. 
About each political issue $m$, an agent $i$ has an opinion $o_{m}^{i}(t)$, which can change in discrete time steps $t=1,...,T$. 
These opinions shall be expressed as real numbers that can be normalized such that they always fall into the interval $[-1,+1]$.
$o_{m}^{i} = -1$ corresponds to strong opposition of $i$ to policy issue $m$, $o_{m}^{i} = 1$ to strong approval, and $o_{m}^{i} = 0$ to a neutral position.  
The political position of each agent $i$ in this multi-dimensional opinion space is characterized by an \emph{opinion vector} $\mathbf{o}^{i}(t)$ composed of the $M$ opinions $o_{m}^{i}(t)$.
Considering a multi-agent system with $i=1,...,N$ agents, the multi-dimensional opinion space is populated with $N$ opinion vectors $\mathbf{o}^{i}$.

Each agent is further characterized by a \emph{affective level} $e^{i}$.
This scalar value does not change over time and can be expressed as a real number from  the interval $[0,1]$.
It describes the level of \emph{affective} involvement of agent $i$ into policy issues. 
$e^{i} = 0$ corresponding to extremely weak, and $e^{i} = 1$ to extremely strong emotional involvement.

\paragraph{Initial setup. \ }

To determine the initial state of these variables, for each agent and each opinion component $o_{m}^{i}$ a random number is sampled from a normal distribution, $\mathcal{N}(\mu_{o},\sigma_{o})$ truncated to the interval $[-1,+1]$.
The mean of the initial opinion components is given as $\mu_{o}(t=0) = 0$ and their standard deviation as $\sigma_{o}(t=0) < 1$.
This ensures that (i) all possible values have indeed a non-negligible probability to occure, but (ii) different from a uniform distribution extreme opinions will not occur with the same probability as moderate opinions, but less frequent.
Thus, the $M$-dimensional opinion space is initially populated with $N$ opinion vectors $\mathbf{o}^{i}(0)$.

Similarly, for each agent the affective level $e^{i}$ is drawn from a truncated normal distribution $\mathcal{N}(\mu_{e},\sigma_{e})$, limited between $0$ and $1$, with mean $\mu_e$ and standard deviation $\sigma_e$.

\paragraph{General opinion dynamics. \ }

To specify the dynamics of the individual opinion vectors, we use the concept of \emph{Brownian agents} \citep{schweitzer2007brownian}.
This means that the dynamics results from an additive superposition of \emph{deterministic} and \emph{stochastic} influences:
\begin{equation}
\mathbf{o}^{i}(t+1) = \mathcal{F}\left[\mathbf{o}^{i}(t),\mathbf{o}^{j}(t)\right]\ \mathcal{G}\left[\mathbf{o}^{i}(t),\mathbf{o}^{j}(t)\right] + \mathcal{Z}\left[\mathbf{o}^{i}(t)\right]
\label{eq:3}
\end{equation}
The deterministic term is composed by two functions, $\mathcal{F}[\cdot]$ and $\mathcal{G}[\cdot]$ that depend on the opinion vector of the focal agent $i$, but also the opinion vectors of other agents $j$, $i$ could potentially interact with.
Specifically, we consider \emph{asynchronous updating} of the dynamics.
That means, at every time step $t$ two agents $i$ and $j$ are selected uniformly at random from the agent pool of $N$ agents.
The term $\mathcal{F}\left[\mathbf{o}^{i}(t),\mathbf{o}^{j}(t)\right]$ then determines whether $i$ and $j$ will interact at all.
This might not be the case if, for example, their opinion vectors $\mathbf{o}_t^{i}$ and $\mathbf{o}_t^{j}$ diverge too much, as we will discuss below.
If $i$ and $j$ interact, then the term $\mathcal{G}\left[\mathbf{o}^{i}(t),\mathbf{o}^{j}(t)\right]$ determines how the opinion vector of $i$ will change based on the influence from $j$. 

The stochastic term, $\mathcal{Z}\left[\mathbf{o}^{i}(t)\right]$, represents \emph{random influences} on the opinion vector of agent $i$, specifically  those influences that do not originate from interactions with other individuals.
For example, an individual's own thought processes may cause it to modify its opinions on various issues without external influences.

\subsection{Considering the Euclidean distance between opinions}
\label{sec:simpl-probl}

\paragraph{Bounded confidence model. \ }

Before we will discuss different functional forms of $\mathcal{F}[\cdot]$, $\mathcal{G}[\cdot]$ and $\mathcal{Z}[\cdot]$, we illustrate the dynamics by turning to the most simple case of a \emph{one-dimensional} opinion space.
Because there is only one policy issue, $M=1$, each agent $i$ only has the opinion $o^{i}(t)\in\{-1,+1\}$
Using the linear transformation $x^{i}= (o^{i}+1)/2$, we can map these opinions to an interval $x^{i}\in\{0,1\}$.
For the dynamics of continuous opinions $x^{i}(t)$ the bounded confidence model was proposed \citep{lorenz2007continuous}.
It assumes that two agents $i$ and $j$ will only interact if the difference between their opinions is smaller than a threshold value $\epsilon$, denoted as the confidence interval, i.e. if the variable $z^{ij}(t)$ is larger than zero: 
\begin{align}
  \label{eq:4}
 z^{ij}(t)= \epsilon - \Delta x^{ij}(t) \geq 0\;; \quad \Delta x^{ij}(t) = \abs{x^{j}(t)-x^{i}(t)} 
\end{align}
We note that, for the one-dimensional case, $\Delta x^{ij}$ gives the \emph{Euclidean distance} between the two opinions.
If the two agents interact, both change their opinions toward the common mean, i.e.
\begin{align}
  \label{eq:5}
  x^{i}(t+1) &= x^{i}(t)+\omega \left[x^{j}(t)-x^{i}(t)\right]\  \Theta\left[z^{ij}(t)\right] \nonumber \\
   x^{j}(t+1) &= x^{j}(t)+\omega \left[x^{i}(t)-x^{j}(t)\right]\  \Theta\left[z^{ji}(t)\right]
\end{align}
Here $\Theta[x]$ is the Heaviside function that  gives $\Theta[x]=1$ if $x\geq 0$ and $\Theta[x]=0$ otherwise.
$\omega$ is the ``speed''  of this change.
If $\omega=0.5$, both agents immediately converge to the mean of their two opinions, i.e. $x^{i}(t+1)=x^{j}(t+1)=[x^{j}(t)+x^{i}(t)]/2$.

Whether or not the multi-agent system converges to a single opinion, denoted as \emph{consensus}, depends on the value of $\epsilon$.
For $\epsilon=0.5$, consensus is obtained, for $\epsilon=0.2$ instead two agent groups with distant opinions emerge.  
The smaller $\epsilon$, the more different opinions coexist in equilibrium \citep{lorenz2007continuous,hk}.
Various extensions of the bounded confidence model have been proposed \citep{DEFFUANT_2018,flache2017models,Pfitzner2013}, also in combination with network dynamics \citep{groeber2009,Weisbuch_2004,Meng_2018,Kurmyshev_2011}.
For the bigger picture of this type of dynamics see also \citep{Schweitzer2020}. 

The bounded confidence model assumes a deterministic dynamics.
Hence, it is expressed by the general opinion dynamics of Eq.~\eqref{eq:3}, if we choose the different functions as follows (with $\mathbf{o}^{i}=o^{i}$ for the one-dimensional case):
\begin{align}
  \label{eq:6}
  \mathcal{F}\left[{o}^{i}(t),{o}^{j}(t)\right]& = \Theta\left[2\epsilon -\abs{o^{j}(t)-o^{i}(t)} \right] \nonumber \\
    \mathcal{G}\left[{o}^{i}(t),{o}^{j}(t)\right] &= o^{i}(t) + \omega \left[o^{j}(t)-o^{i}(t)\right] \nonumber \\
   \mathcal{Z}\left[{o}^{i}(t)\right] &=  0
\end{align}

\paragraph{Two-dimensional opinions. \ }

\begin{figure}[htbp]
  \centering
  \includegraphics[width=0.5\textwidth]{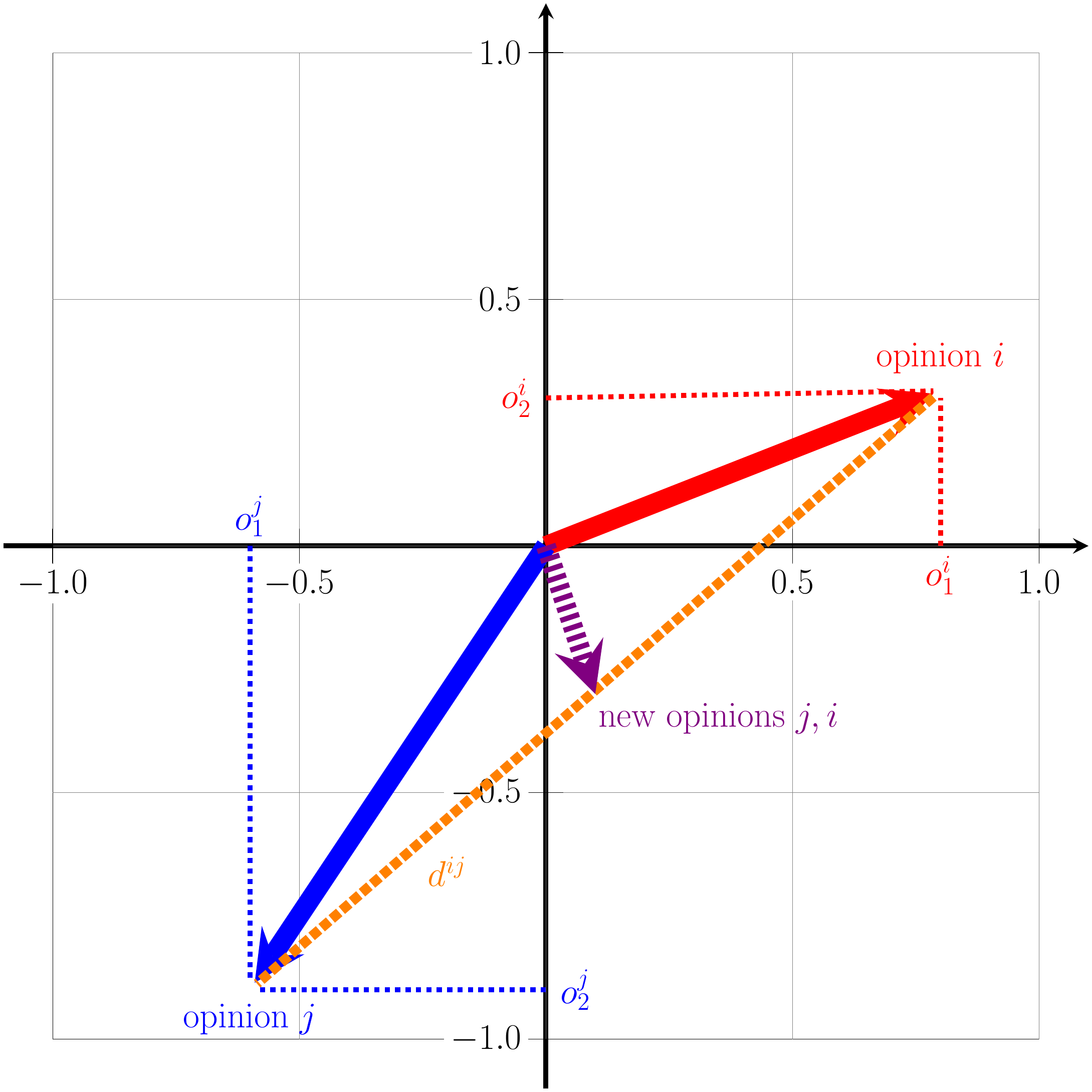}
  \caption{Opinion vectors of two agents $i$ and $j$ in a two-dimensional opinion space. $d^{ij}$ denotes the Euclidean distance, Eq. \eqref{eq:7}. The new opinion for both agents follows from Eq. \eqref{eq:6}.}
  \label{fig:2d}
\end{figure}

The bounded confidence model can be formally generalized towards multi-dimensional problems, but the results are \emph{not} trivial, as we will show. 
As an illustration, we first use a two-dimensional opinion space shown in Figure~\ref{fig:2d}. 
There, each agent $i$ has an opinion about issues 1 and 2, denoted by the opinon vector $\mathbf{o}^{i}(t)=o^{i}_{1}(t)\mathbf{s}_{1}+o^{i}_{2}(t)\mathbf{s}_{2} \equiv \{o_{1}^{i}(t),o_{2}^{i}(t)\}$.
Here $\mathbf{s}_{1}$, $\mathbf{s}_{2}$ denote the unit vectors (versors) of the respective coordinate axes. 
That means, in order to decide whether two agents interact, we have to determine the \emph{similarity} of their \emph{opinion vectors} at a given time step $t$. 

One possible measure to determine the difference between the two opinion vectors is the \emph{Euclidean distance} 
$d^{ij}(t)=d\left[\mathbf{o}^{i}(t),\mathbf{o}^{j}(t)\right]$, which is defined for the two-dimensional opinion space as:
\begin{equation}
d^{ij}(t)= \sqrt{\left[o^{i}_{1}(t)-o^{j}_{1}(t)\right]^{2}+\left[o^{i}_{2}(t)-o^{j}_{2}(t)\right]^{2}}
\label{eq:7}
\end{equation}
Indeed, the Euclidean distance is applied by 
political scientists to measure the similarity in opinion space between two political actors.
Greater distance then corresponds to less similarity \citep{benoit2006party}.
To normalize the Euclidean distance to values between $0$ and $1$, it has to be divided by the length of the diagonal of the opinion space, $\sqrt{4M}=2\sqrt{2}$. 

We now assume as in the bounded confidence model that two agents $i$ and $j$ interact if their normalized Euclidean distance is less than a given threshold value $2\epsilon$.
As a result of this interaction, both agents adjust their opinions \emph{component wise} to the common mean.
With $\omega=0.5$, we choose for the two-dimensional case the different functions in the general opinion dynamics of Eq. \eqref{eq:3} as follows:  
\begin{align}
  \label{eq:6aa}
  \mathcal{F}\left[\mathbf{o}^{i}(t),\mathbf{o}^{j}(t)\right]& = \Theta\left[4\sqrt{2}\,\epsilon - d^{ij}(t)\right] \nonumber \\
  \mathcal{G}\left[\mathbf{o}^{i}(t),\mathbf{o}^{j}(t)\right] &= \frac{\mathbf{o}^{i}(t) + \mathbf{o}^{j}(t)}{2}
 =  \frac{{o}_{1}^{i}(t) + {o}_{1}^{j}(t)}{2} + \frac{{o}_{2}^{i}(t) +{o}_{2}^{j}(t)}{2} \nonumber \\
   \mathcal{Z}\left[\mathbf{o}^{i}(t)\right] &=  \mathbf{\xi}_{i}(t)
\end{align}
At difference with the deterministic bounded confidence model, here we have added a random vector drawn from a truncated normal distribution, limited between $-1$ and $1$, with mean $\mu_\xi = 0$ and standard deviation $0 < \sigma_\xi < 1$.
This shall account for stochastic influences on the opinion formation not related to the interactions.

\subsection{Results of agent-based simulations}
\label{sec:model1}

\begin{figure}[htbp]
  \includegraphics[width=.29\linewidth]{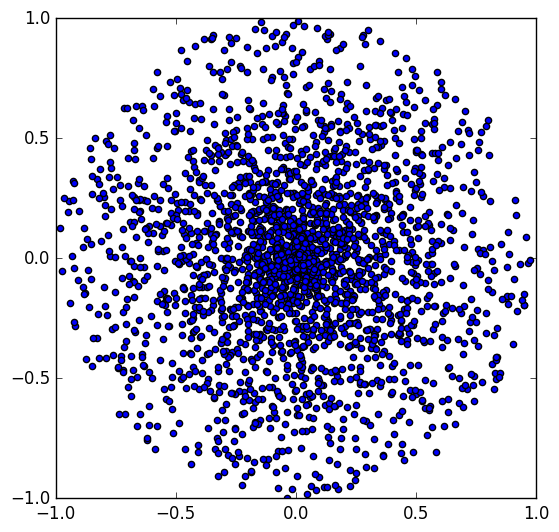}(a)
  \hfill
  \includegraphics[width=.29\linewidth]{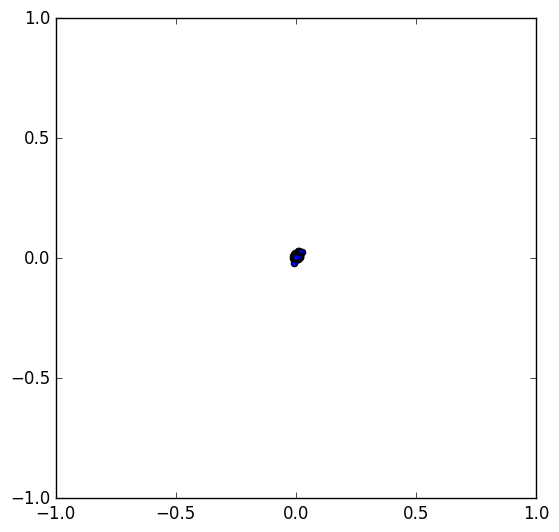}(b)
  \hfill
\includegraphics[width=.29\linewidth]{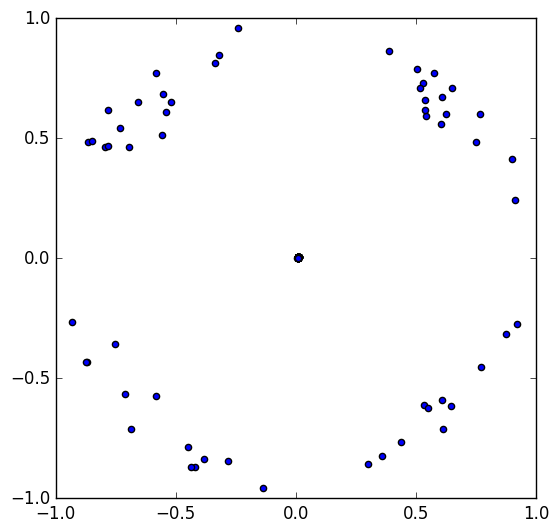}(c)
\caption{Opinions of $N$=10.000 agents in a two-dimensional opinion space. (a) Initial distribution at time $t=0$. (b) Long-term outcome ($t=60.000$) if $\epsilon=0.5$ . (c) Long-term outcome ($t=210.000$) if $\epsilon=0.25$. Other parameters see Appendix \ref{sec:simul-two-dimens}: 
}
  \label{fig:2d-bounded}
\end{figure}

We illustrate the dynamics of the two-dimensional bounded confidence model, Eq. \eqref{eq:6aa}, by means of stochastic simulations of the multi-agent system.
The final results are shows in Figure~\ref{fig:2d-bounded} for two different values of the bounded confidence interval $\epsilon$.
The details of the dynamics are presented in Figure~\ref{fig:evolution} in Appendix~\ref{sec:simul-two-dimens}. 

Figure~\ref{fig:2d-bounded}(a) shows the initial state for our simulations, where agents got randomly assigned an opinion vector in the two-dimensional opinion space, as described in Section~\ref{sec:agent-based-model}.
Figure~\ref{fig:2d-bounded}(b) shows the outcome of the opinion dynamics if a rather large confidence interval $\epsilon$ is chosen, while Figure~\ref{fig:2d-bounded}(c) shows the outcome for a rather small value of $\epsilon$.
The results are in line with insights from the one-dimensional bounded confidence model, Eq. \eqref{eq:5}.
If $\epsilon$ is large enough, we see the emergence of \emph{consensus} in the middle of the opinion space.
For the classical bounded confidence model, this would be $x^{\mathrm{stat}}=0.5$, whereas it is here $\{o_{1}^{\mathrm{stat}},o_{2}^{\mathrm{stat}}\}=\{0,0\}$.
If $\epsilon$ is too small to reach consensus, we observe the formation of different clusters in opinion space, i.e. groups of agents converging to the same opinion.
This indicates the long-term \emph{coexistence} of different opinions in the multi-agent system.
We note that, in this case, still the \emph{majority} of vectors converge to the origin, which shows the largest cluster of agents, whereas  in the periphery, some agents are left behind in smaller clusters. 

While the outcome reached is mainly determined by the value of $\epsilon$, it also depends on the level of randomness, expressed by $\xi$.
If the standard deviation $\sigma_{\xi}$ is sufficiently large, random changes of opinions are able to bring agents sufficiently close 
in opinion space such that they can continue to interact.
This then fosters the emergence of consensus, by destabilizing opinion clusters in the periphery.
Whether the outcome of the simulations results in consensus or coexistence is certainly different from the agent perspective, but the resulting \emph{average opinion} over all agents in the long-run is in both cases the \emph{same}, namely $\{\bar{o}_{1},\bar{o}_{2}\}=\{0,0\}$.

\section{Modeling multi-dimensional opinion alignment}
\label{sec:alig}

\subsection{Proximity voting vs. directional voting}
\label{sec:subj-simil-opin}

The above simplified model of multi-dimensional opinion dynamics assumes that the Euclidean distance in opinion space is a valid proxy to measure the similarity between the opinion vectors of two agents.
While this distance can be calculated, its real meaning becomes questionable if we expand the opinion space to higher dimensions $M$.
Moreover, as Eq. \eqref{eq:7} shows, the opinions regarding different issues $m$ are treated as independent, thus differences between opinions can be simply added component wise.
This assumption is not undisputed because of its rather weak empirical evidence  \citep{benoit2006party}.
It was argued that its widespread usage in political science is rather due to its convenient mathematical properties.

In explanations of voting behavior the assumption to use the Euclidean distance as a similarity measures is called \emph{proximity voting}.
It is challenged by a different paradigm, called \emph{directional voting} \citep{rabinowitz1989directional}.
This means, voters do not vote for the candidate that is closest to them in opinion space, but for the candidate which is on the 'right side' of most issues.
This argument was already explained in relation to Figure~\ref{fig:struct_bal}(a). 

We can illustrate the difference between the two paradigms with a simple calculation.
Let us consider three agents with the following opinion vectors in a two-dimensional policy space, $\mathbf{o}^{i} = (0.1,0.1)$, $\mathbf{o}^{j} = (0.3,0.3)$, and $\mathbf{o}^{k} = (-0.1,-0.1)$.
From Eq.~\eqref{eq:7} we know the Euclidean distances with respect to agent $i$, i.e. $d^{ij}=\sqrt{0.08}$ and $d^{ik}=\sqrt{0.08}$.
Thus, if the perception of similarity is based on proximity in Euclidean space, $i$ would perceive the opinion of $j$ exactly as similar as the opinion of $k$, because it has the same Euclidean distance to both.
But is this assumption realistic?
Certainly not, if positive values on a given policy issue, e.g. marijuana legalization, represent a positive stance, and negative values a negative stance. 
Both $i$ and $j$ are on the 'pro' side of both policy dimensions, $o^{i}_{1}>0$,  $o^{i}_{2}>0$,  $o^{j}_{1}>0$,   $o^{j}_{2}>0$, while $k$ is on the 'contra' side,  $o^{k}_{1}<0$,  $o^{k}_{2}<0$.
The difference between $i$ and $j$ only lies in the strength of their approval to both issues.
From this perspective, it would make sense if $i$ and $j$ would perceive each other as very similar and $k$ as very dissimilar.
This is exactly what the \emph{directional voting} paradigm postulates.
The question whether it  is more realistic is still debated \citep{lacy2010testing}.
But in the following, we will implement \emph{directional} voting in our multi-dimensional opinion dynamics model, to contrast the results with \emph{proximity} voting, shown in Section~\ref{sec:simpl-probl} and Appendix \ref{sec:appendix}. 
That means, we will propose new similarity measures that enter the function $\mathcal{F}\left[\mathbf{o}^{i}(t),\mathbf{o}^{j}(t)\right]$.
But we will also test a different assumption of how agents respond to the opinions of others, expressed in $\mathcal{G}\left[\mathbf{o}^{i}(t),\mathbf{o}^{j}(t)\right]$.

\subsection{Considering directional similarity}

\paragraph{Transformation to polar coordinates. \ }

In order to apply the directional voting paradigm, we have to modify the way agents quantify distances between opinions.
For this, we introduce a new measure, \emph{directional similarity}, $D^{ij}$.
While the Euclidean distance $d^{ij}$, Eq. \eqref{eq:7}, takes the full information from the opinion vectors $\mathbf{o}^{i}$ and $\mathbf{o}^{j}$ into account, the directional similarity only uses information about the \emph{angles} $\phi^{i}$ and $\phi^{j}$ of the respective opinion vectors.

To formalize this step, we transform the opinion vector $\mathbf{o}^{i}(t)=\left\{o^{i}_{1}(t),o^{i}_{2}(t)\right\}$ into polar coordinates,  $\mathbf{o}^{i}(t)=\left\{\abs{o^{i}(t)},\phi^{i}(t)\right\}$, where the length $\abs{o^{i}(t)}$ and the angle $\phi^{i}(t)$ of the vector are, for a two-dimensional opinion space, defined as follows:
\begin{align}
  \label{eq:8}
  \abs{o^{i}(t)}&= \sqrt{\left[o^{i}_{1}(t)\right]^{2}+\left[o^{i}_{2}(t)\right]^{2}} \nonumber \\
  \phi^{i}(t)& = \left\{
\begin{array}{ll}
        \arctan{\left[o^{i}_{2}(t)/o^{i}_{1}(t)\right]}, & \text{for } o^{i}_{1}(t) >0, o^{i}_{2}(t) \geq 0\\
  \arctan{\left[o^{i}_{2}(t)/o^{i}_{1}(t)\right]} + 2\pi, & \text{for } o^{i}_{1}(t) > 0, o^{i}_{2}(t) < 0\\
  \arctan{\left[o^{i}_{2}(t)/o^{i}_{1}(t)\right]} + \pi, & \text{for } o^{i}_{1}(t) < 0
        \end{array}
               \right.
\end{align}
The case analysis is needed because $\arctan(x)$ is not an injective function, but it is convenient to implement.  
This always returns a value $\phi\in[0,2\pi]$.
We can then define the difference between the angles of the opinion vectors of agents $i$ and $j$ as:
\begin{align}
  \label{eq:9}
   \Delta \phi^{ij}(t)& = \left\{
\begin{array}{rl}
        \left[\phi^{j}(t)-\phi^{i}(t)\right], & \text{if }  \abs{\phi^{j}(t)-\phi^{i}(t)} \leq \pi \\
  2\pi -  \left[\phi^{j}(t)-\phi^{i}(t)\right], & \text{if } \abs{\phi^{j}(t)-\phi^{i}(t)} > \pi
        \end{array}
               \right.
\end{align}
This always returns a value $\Delta\phi\in[0,\pi]$, which can be mapped to an interval $[0,1]$ by scaling $\Delta\phi/\pi$.

\paragraph{Opinion dynamics based on opinion alignment. \ }

To specify the general opinion dynamics, Eq.~\eqref{eq:3}, we make the following assumption for $\mathcal{F}\left\{\mathbf{o}^{i}(t),\mathbf{o}^{j}(t)\right\}$: two randomly chosen agents $i$ and $j$ will only interact if $\Delta \phi^{ij}(t)/\pi$ is less than a critical threshold $\epsilon^{i}$.
Different from the bounded confidence model the value of $\epsilon$ now becomes an \emph{individual} parameter.
Specifically, we assume that it is coupled to the \emph{affective involvement} $e^{i}$.
As mentioned above, agents with a high level of affective involvement may become less tolerant to other opinions.
Therefore, we define $\epsilon\equiv \epsilon^{i}=1-e^{i}$, where $e^{i}$ initially is randomly chosen from the interval $[0,1]$ and constant over time.
This results in
\begin{align}
  \label{eq:10}
  \mathcal{F}\left\{\mathbf{o}^{i}(t),\mathbf{o}^{j}(t)\right\} = \Theta\left[D^{ij}(t) -e^{i} \right] \;; \quad
  D^{ij}(t)= 1- \frac{\Delta \phi^{ij}(t)}{\pi}
\end{align}
We call $D^{ij}(t)$ the \emph{pairwise directional similarity}.
It becomes maximal, $D^{ij}=1$, if both agents have perfectly aligned opinion vectors.
In this case even a maximal affective involvement, i.e. a minimal confidence interval, will not prevent them from interacting. 

\begin{figure}[htbp]
  \centering
  \includegraphics[width=0.45\textwidth]{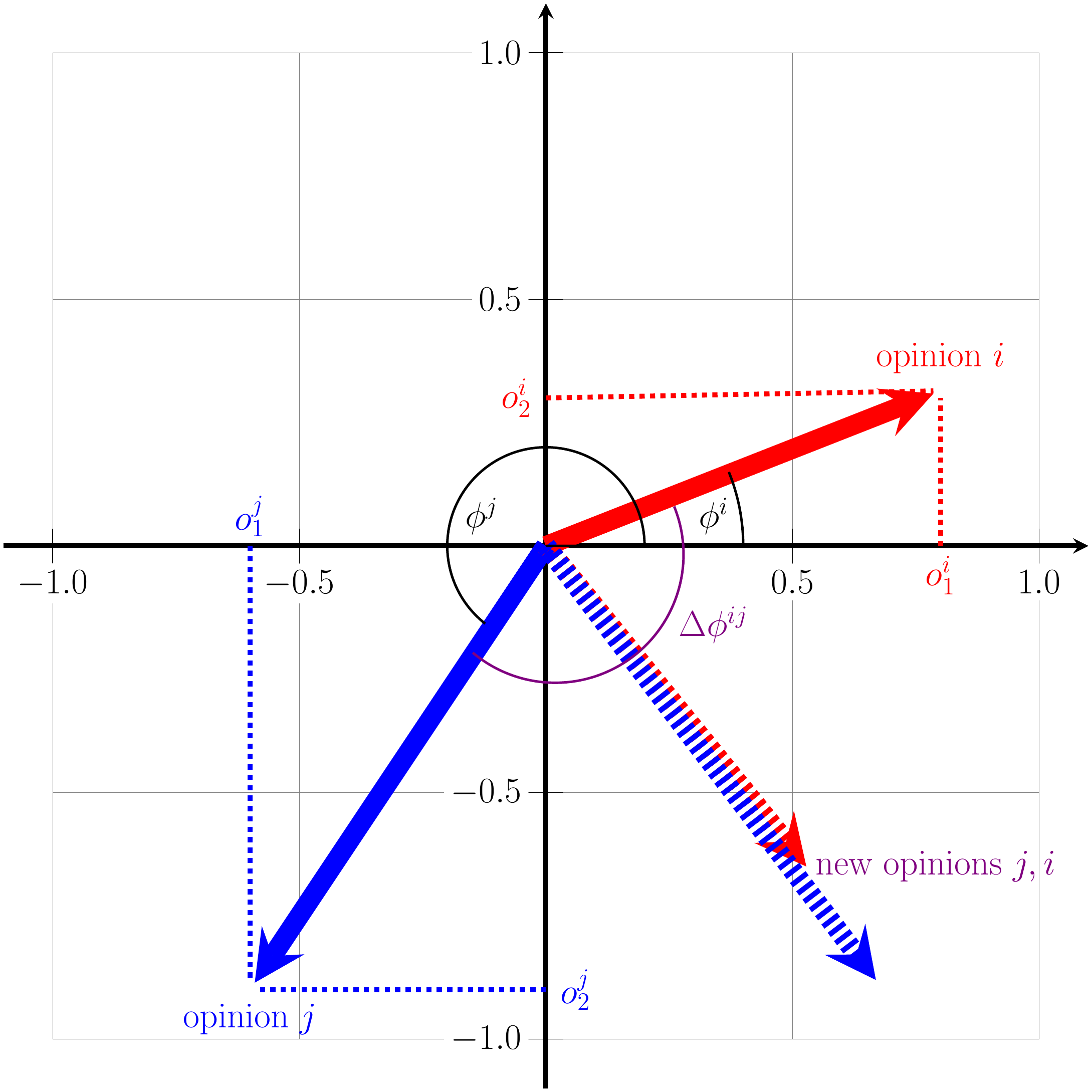}
  \caption{Opinion vectors of two agents $i$ and $j$ in a two-dimensional opinion space. $\Delta \phi^{ij}$ is given by Eq.~\eqref{eq:9}. The new opinion for both agents follows from Eq.~\eqref{eq:13}. }
  \label{fig:three}
\end{figure}

If the two agents interact, then they change their opinion such that they align their opinion vectors.
I.e. the opinion vectors \emph{rotate} to a new angle $\phi^{i}\to \theta^{i}$, but their absolute value does \emph{not} change.
The update function $\mathcal{G}\left\{\mathbf{o}^{i}(t),\mathbf{o}^{j}(t)\right\}$ therefore reads in Cartesian coordinates as:
\begin{align}
  \label{eq:11}
  \mathcal{G}\left\{\mathbf{o}^{i}(t),\mathbf{o}^{j}(t)\right\}= \abs{\mathbf{o}^{i}(t)}
\left[ \cos{\left\{\theta^{i}(t)\right\}}\mathbf{s}_{1}+\sin{\left\{\theta^{i}(t)\right\}}\mathbf{s}_{2}\right]
\end{align}
Random influences now only affect the angle $\phi^{i}$, i.e.
\begin{align}
  \label{eq:14}
  \mathcal{Z}\left[\mathbf{o}^{i}(t)\right]= \xi(t)\phi^{i}(t)
\end{align}
Therefore, the updated angle $\theta^{i}(t)$ is determined both by the (deterministic) rotation and by random changes $\xi(t)$:
\begin{align}
  \label{eq:12}
\theta^{i}(t) = \phi^{i}(t) \left[1+\xi(t)\right] + \omega \Delta \phi^{ij}(t) 
\end{align}
If we assume as before $\omega=0.5$, we find in explicit form:
\begin{align}
  \label{eq:13}
  \theta^{i}(t) = \xi(t)\phi^{i}(t) + \left\{
\begin{array}{rl}
        \frac{\displaystyle \left[\phi^{j}(t)+\phi^{i}(t)\right]}{\displaystyle 2}, & \text{if }  \abs{\phi^{j}(t)-\phi^{i}(t)} \leq \pi \\
  \pi + \frac{\displaystyle \left[\phi^{j}(t)+\phi^{i}(t)\right]}{\displaystyle 2}, & \text{if } \abs{\phi^{j}(t)-\phi^{i}(t)} > \pi
        \end{array}
               \right.
\end{align}
This update rule is illustrated in Figure~\ref{fig:three}, to be compared to Figure~\ref{fig:2d} based on the Euclidean distance. 

\subsection{Results of agent-based simulations}
\label{sec:results-agent-based}

\begin{figure}[htbp]
  \includegraphics[width=.29\linewidth]{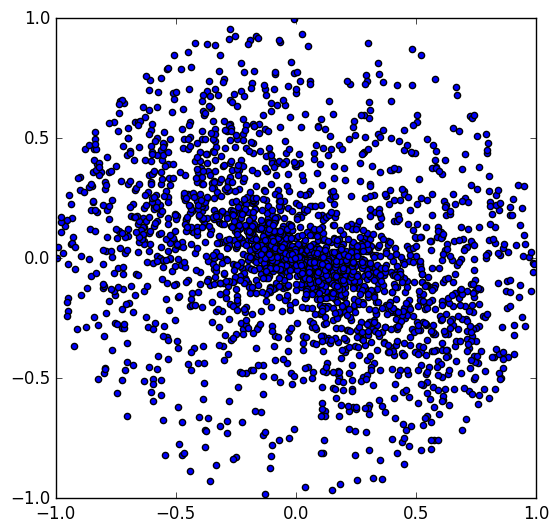}(a)
  \hfill
  \includegraphics[width=.29\linewidth]{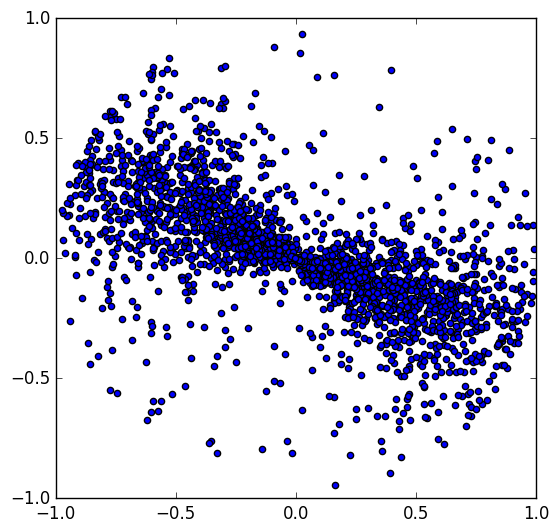}(b)
  \hfill
\includegraphics[width=.29\linewidth]{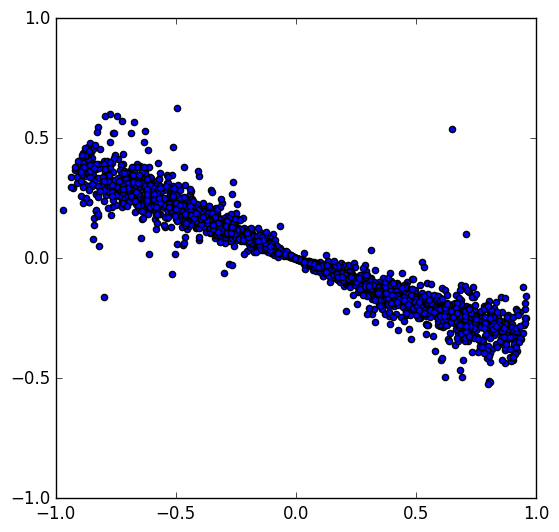}(c)

\includegraphics[width=.29\linewidth]{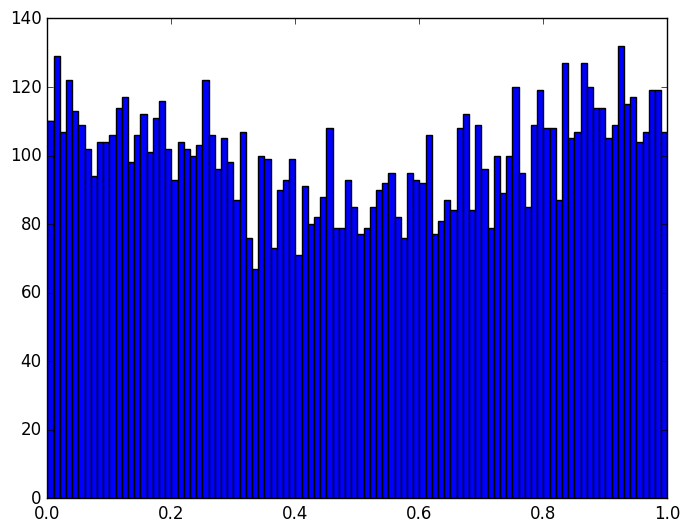}
\hfill
\includegraphics[width=.29\linewidth]{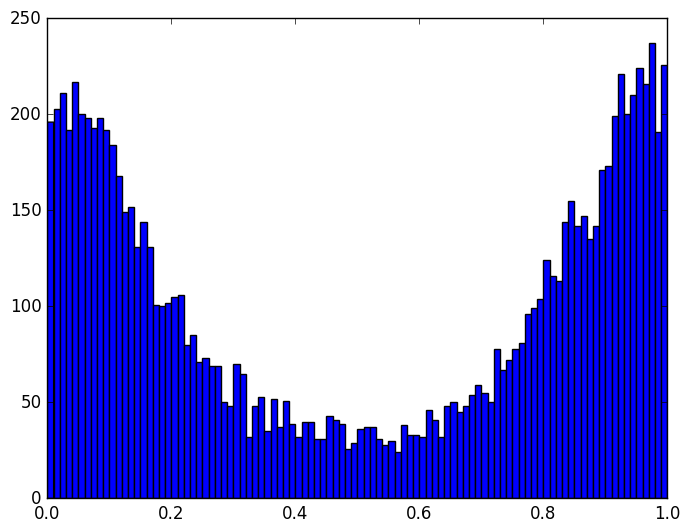}
\hfill
\includegraphics[width=.29\linewidth]{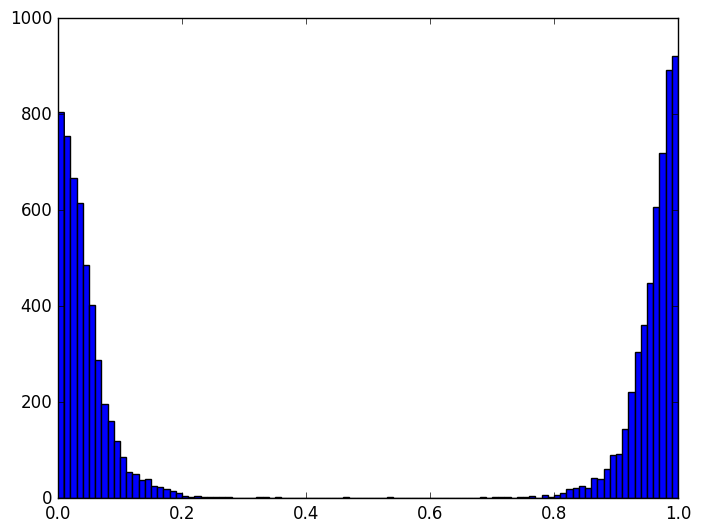}$\quad$
\caption{(Top row) Opinions of $N=2500$ agents in a two-dimensional opinion space at different time steps: (a) $t=50.000$, (b) $t=70.000$, (c) $t=100.000$.
 (Bottom row) Distribution of the corresponding pairwise directional similarity $P[D^{ij}(t)]$, Eq.~\eqref{eq:10}.  }\label{fig:aligned_polarization}
\end{figure}

We illustrate the outcome of the alignment model by means of agent-based simulations illustrated in Figure~\ref{fig:aligned_polarization}.
To make it comparable to Figure~\ref{fig:2d-bounded}, we first restrict ourselves to the two-dimensional opinion space. 
The initial state is the same as shown in Figure~\ref{fig:2d-bounded}(a) and follows from the setup described in Section~\ref{sec:agent-based-model}.
While the first row in Figure~\ref{fig:aligned_polarization} shows the positions of the agents in the two-dimensional opinion space at different times, 
the second row shows the distribution of corresponding pairwise similarity measure, $D^{ij}(t)$, Eq.~\eqref{eq:10}.
The initial distribution of $D^{ij}(0)$, which matches the initial opinion distribution, Figure~\ref{fig:2d}(a), is shown in Figure~\ref{fig:initial_distribs}(a).

In the simulations shown in Figure~\ref{fig:aligned_polarization}, the affective involvement $e^{i}$, which together with $D^{ij}$ enters the function $\mathcal{F}[\cdot]$, Eq.~\eqref{eq:10}, is set to a constant value $e^{i}\equiv e =0.5$, equal for all agents.
Eventually, random influences are set to zero, $\mathcal{Z}[\cdot]=0$, Eq.~\eqref{eq:14}, i.e., the dynamics are completely deterministic. 

As we see, in an early phase the opinion vectors are broadly distributed, and the corresponding distribution $P[D^{ij}(t)]$, which reflects the angle $\Delta \phi^{ij}(t)$ between any two vectors, is almost uniform.
This changes over time into a clear bimodal distribution.
Its meaning becomes clear from the opinion positions in the two-dimensional space: agents tend to align their opinions such that a dominant direction emerges.
Almost all agents align to this dominant direction, but still position themselves on opposite sites.
Hence, we do \emph{not} observe \emph{consensus} (which would also imply an alignment of opinions), but the \emph{coexistence} of opinions from the left/right spectrum, i.e. \emph{polarization}.  

Note that, because of the assumed directional voting, agents do not adjust the magnitude of their opinion vectors, but just the angle.
We add that a recent model built on a novel \emph{weighted balance theory} \citep{schweighofer2019} is also able to reflect changes in the magnitude of the opinion vectors. 
Without that, we do not see a pronounced polarization, in which extreme opinions (with large magnitude) dominate.
But the emergence of a \emph{global alignment} is clearly observed, which was the goal of this opinion dynamics model. 

\subsection{Higher-dimensional opinion space}
\label{sec:high-dimens-opin}

While our opinion dynamics model performs very promising in two dimensions, this raises the question how robust
the outcome is if we change (i) interaction parameters, or (ii) the dimensionality of the opinion space.
Unfortunately, this robustness is not given, and in the following we shortly explain the reasons for this, as a motivation for the model extension in Section~\ref{sec:abm_3}. 

The underlying dynamics assumes a critical threshold, in this case expressed by the affective involvement $e$.
Very similar to the bounded confidence model, this threshold decides about \emph{consensus} or \emph{coexistence} of opinions in the long run.
Decreasing $e$ for all agents increases the range of possible interactions, because agents with a lower affective involvement have a larger tendency to change their opinions.
This in turn destroys the coexistence of different opinions and fosters consensus.
The same happens if we, instead of a fixed value for all agents, increase the \emph{width} of the distribution, $P[e]$, this way allowing more agents to have a lower value of $e^{i}$.
Of course, there will be also more agents with large $e^{i}$, but what matters to reach consensus is the fraction of those agents that can still interact with others.
If agent $i$ is no longer willing to approach agent $j$ in opinion space, agent $j$ still can if its $e^{j}$ value is low enough.
Eventually, increasing the randomness in the dynamics by setting $\sigma_{\xi}>0$ also favors the emergence of \emph{consensus}.

  \begin{figure}[htbp]
    \centering
\includegraphics[width=0.6\textwidth]{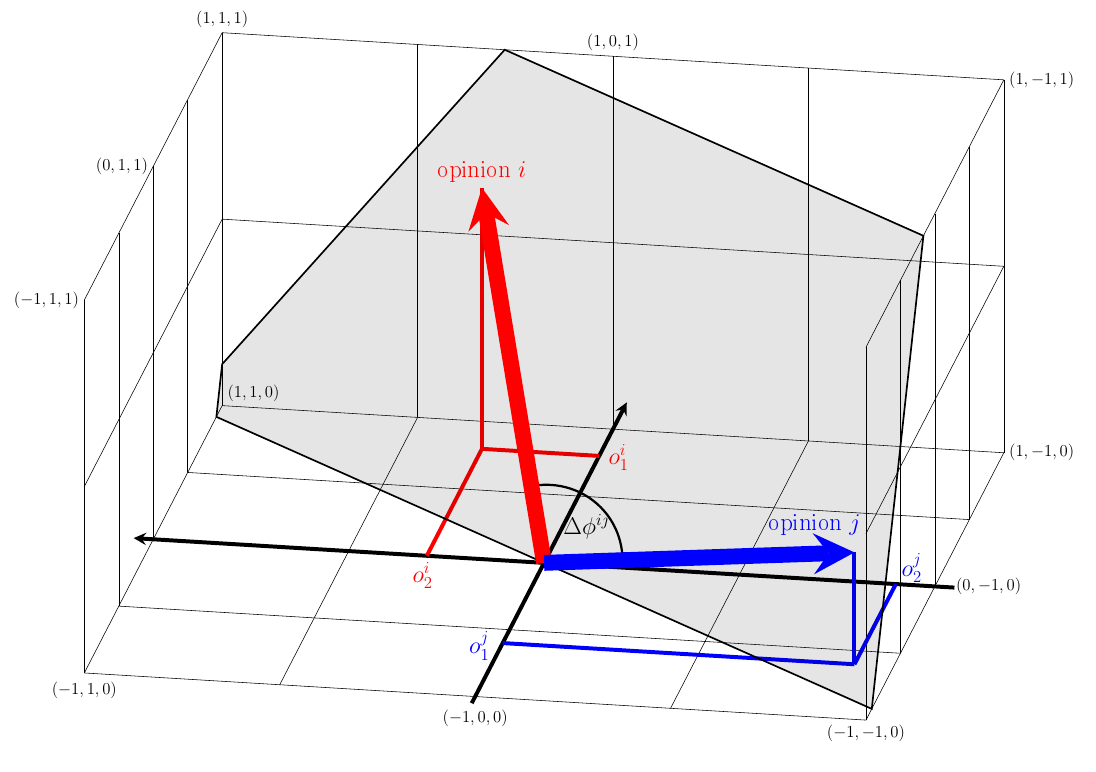}
    \caption{Opinion vectors of two agents $i$ and $j$ in a three-dimensional opinion space. Note that the 2 vectors define a plane (in gray), on which the angle $\Delta \phi^{ij}$ is measured.}
    \label{fig:3d}
  \end{figure}

\begin{figure}[htbp]
  \includegraphics[width=0.29\textwidth]{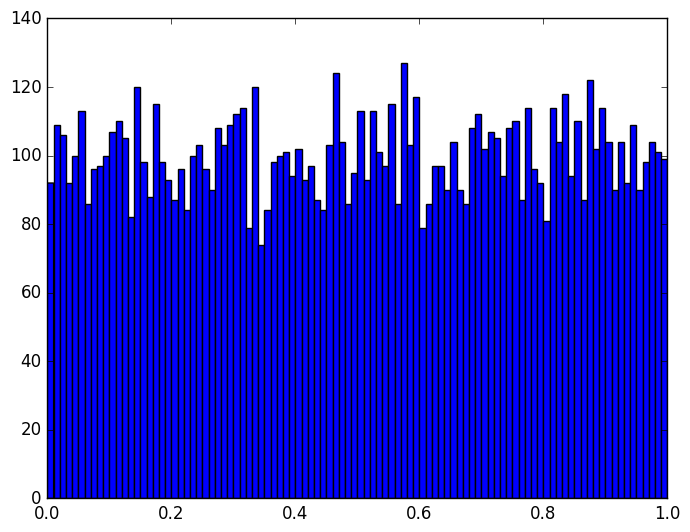}(a)
  \hfill
  \includegraphics[width=0.29\textwidth]{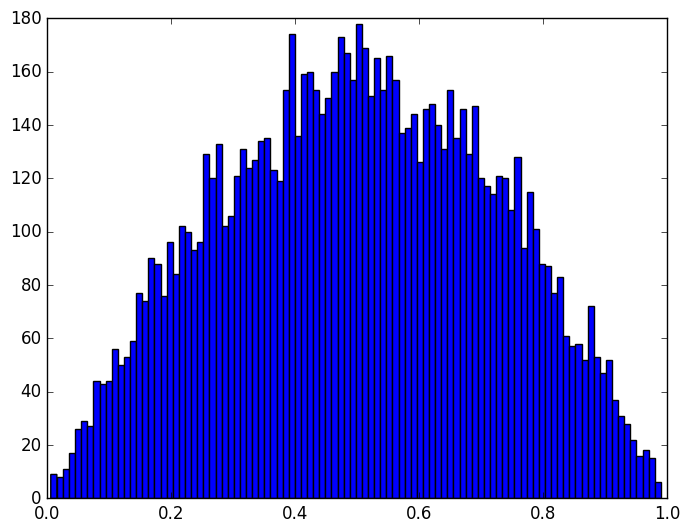}(b)
  \hfill
        \includegraphics[width=0.29\textwidth]{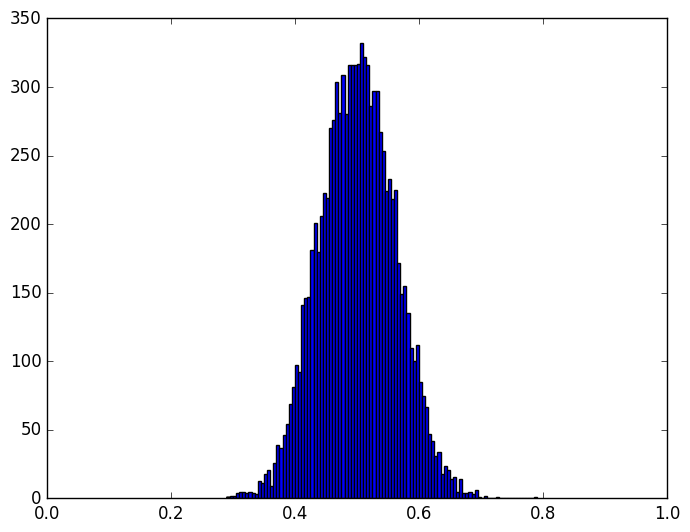}(c)
    \caption{Initial distributions of the pairwise directional similarity $P[D^{ij}(0)]$ for different dimensions of the opinion space: (a) $M=2$, (b) $M=3$, (c) $M=28$. }
    \label{fig:initial_distribs}
  \end{figure}

The dimensionality $M$ of the opinion space impacts the results in a less obvious, but interpretable manner.
We recall that agents get assigned initial opinions on each dimension in a random manner.
The angle between any two  opinion vectors is, in a multi-dimensional space, calculated on the \emph{plane} defined by the two vectors (see Figure~\ref{fig:3d}).
That means, even in higher dimensions, there is only \emph{one} angle $\Delta \phi^{ij}$.
Form $M=2$, this can have initially any value between $(0,\pi)$ (shown  in Figure~\ref{fig:initial_distribs}a).
But the expectation value is $\mean{\Delta \phi^{ij}(0)}=\pi/2$. 
With each additional dimension, the probability to still find extreme values for $\Delta \phi^{ij}$ decreases and the distribution $P[D^{ij}(0)]$ narrows down toward the expectation value, $\mean{D^{ij}(0)} = 1- \mean{\Delta \phi^{ij}(0)}/{\pi} = 0.5$.
This can be clearly seen in Figure~\ref{fig:initial_distribs}, which shows the distribution of initial pairwise distances $P[D^{ij}(0)]$ for different dimensions.
While this distribution is very broad and almost uniform for $M=2$, it quickly approaches a unimodal distribution centered around $0.5$, if $M$ is increased.
In other words, it becomes unlikely for an agent to meet another agent with very similar or very dissimilar opinions.
Most pairs of agents will have a mixture of congruent and opposing opinions.
One could reinforce a larger initial dis/similarity by tweaking the initial conditions, e.g. by sampling more extreme opinions with a higher probability.
But this would need some arguments that can hardly be satisfied. 

For higher dimensions, the initial distribution $P[D^{ij}(0)]$ already ensures that all agents have (almost) the same alignment/disalignment of their opinion vectors, i.e. $\pi/2$.
Then, the  threshold value $e$, or the respective distribution $P[e]$, determines the outcome of the opinion dynamics.
If the affective involvement is low, for example $e\ll 0.5$, the majority of agents is able to interact, this way aligning their opinions even more, which eventually leads to a large \emph{alignment} together with (almost perfect) \emph{consensus}.
This is shown in Figure~\ref{fig:28_D} in the Appendix.
If the threshold is high,  $e\gg 0.5$, the majority of agents is not able to interact.
Then their alignment distribution stays as a unimodal distribution centered around 0.5, very close to the initial distribution.
In both cases, it is \emph{not} possible to obtain the desired scenario of a \emph{bimodal} alignment distribution, where agents align their opinions along the emerging ``ideological dimension'', as shown in Figure~\ref{fig:aligned_polarization} for $M=2$. 
The lack of diametrically opposed opinion vectors in the initial state makes the emergence of a polarized state very unlikely.

\section{Modeling multi-dimensional opinion alignment with repulsion}
\label{sec:abm_3}

\subsection{Interactions without constraints}
\label{sec:inter-with-constr}

So far, we have used a critical threshold to determine whether two agents still interact.
This threshold has already become an individual parameter and was coupled to the affective involvement of the agents.
But now we go back to the argumentation in Section~\ref{sec:probl-multi-dimens}, where we discussed the options of two individuals that disagree on most issues and only agree on a few.
In addition to \emph{not interacting}, which was modeled above, we now assume that they still interact, but solve their cognitive dissonances by \emph{disagreeing} even on the few issues they previously agreed on.

Because, now every two agents will always interact, we have to change the respective function as follows:
\begin{align}
  \label{eq:15}
  \mathcal{F}\left[\mathbf{o}^{i}(t),\mathbf{o}^{j}(t)\right]=1
\end{align}
However, based on their interactions, two agents will not always align their opinion vectors.
Instead, if $\Delta \phi^{ij}(t)$ is \emph{larger} than a threshold $\delta$, we assume that as a result of their interaction they deviate even more in opinion space.
Precisely, the absolute distance between their updated angles, $\abs{\theta^{j}(t)-\theta^{i}(t)}$ \emph{increases} compared to  $\Delta \phi^{ij}(t)$ if $\Delta \phi^{ij}(t)>\delta$, whereas it \emph{decreases} if  $\Delta \phi^{ij}(t)<\delta$. 
We set $\delta=\pi/2$, which means that agents with orthogonal opinion vectors do not influence each others opinion.
If their current alignment is less then $90\degree$, they tend to align more, if it is more than $90\degree$, they tend to deviate even more.

This assumption is in line with the arguments in Sections~\ref{sec:comb-cogn-diss} and \ref{sec:incr-disal}, because  $\Delta \phi^{ij}(t)$ measures precisely whether agents agree or disagree on most issues.
If  $\Delta \phi^{ij}(t)<\pi/2$, they agree on most issues (but may still disagree on a few), and if  $\Delta \phi^{ij}(t)> \pi/2$ it is the other way round.
That means  $\Delta \phi^{ij}(t)$ effectively determines whether the relation between the two agents is positive, $r^{ij}=+1$, or negative, $r^{ij}=-1$.
This way, we have implemented the theoretical arguments based on the combination of cognitive dissonance theory and structural balance theory in our agent-based model of opinion dynamics.

\begin{figure}[htbp]
    \centering
        \includegraphics[width=0.48\textwidth]{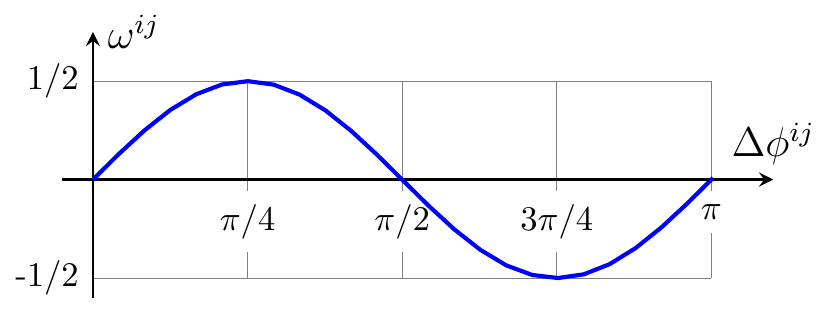}
    \caption{Update parameter $\omega^{ij}=\omega[\Delta \phi^{ij}(t)]$, Eq.~\eqref{eq:16} dependent on the angle $\Delta^{ij}\phi$ between the opinion vectors of agents $i$ and $j$.}
    \label{fig:attract_repulse}
\end{figure}

We can formally express this argument in our update function, $\mathcal{G}\left[\mathbf{o}^{i}(t),\mathbf{o}^{j}(t)\right]$, Eq.~\eqref{eq:11}, if the parameter $\omega$ to update the angle, Eq.~\eqref{eq:12}, becomes a function that depends on $\Delta \phi^{ij}(t)$ in a non-monotonous manner, for example:
\begin{align}
  \label{eq:16}
  \omega^{ij}(t) = \omega\left[\Delta \phi^{ij}(t)\right] = \frac{1}{2}\sin{\left[2\Delta \phi^{ij}(t)\right]}
\end{align}
This is shown in Figure~\ref{fig:attract_repulse}.
As we see, the previous dynamics, i.e. $\omega^{ij}=1/2$, is regained if $\Delta \phi^{ij}(t)=\pi/4$.
In this case, the two agents completely align their opinion, i.e. each one rotates its opinion vector by $\pi/8$ \emph{towards} the other.
Conversely, if $\Delta \phi^{ij}(t)=3\pi/4$, we obtain $\omega^{ij}=-1/2$ and each agent rotates its opinion vector by $\pi/8$ \emph{away} from the other.  

We emphasize that the transition from alignment to disalignment is smooth, and not abrupt.
The largest changes in opinions occur when both the motivation to change opinions and the number of opinions that can be changed are high.
This motivation is high, if agents already have a sufficient agreement or disagreement on a number of issues. 
In cases of perfect alignment or disaligment, agents will not change their opinions based on the interaction with others, because they already agree or disagree completely.

Eventually, we need to specify the function $\mathcal{Z}\left[\mathbf{o}^{i}(t)\right]$ for the random influences.
This still affects the angle $\phi^{i}$, but instead of just considering random shocks $\xi(t)$, we now also consider the affective involvement $e^{i}$ of an agent. 
Assuming that a higher emotional involvement in policy issues makes the opinion of an agent more resistant to \emph{random change}, we choose: 
\begin{align}
  \label{eq:14aa}
  \mathcal{Z}\left[\mathbf{o}^{i}(t)\right]= \phi^{i}(t)\; \xi(t)\left\{1-e^{i}\right\}
\end{align}
That means, agents with higher emotional involvement are less susceptible to noise. 
As before, $\xi$ is sampled from a distribution with mean $\mu_{\xi}=0$ and standard deviation $\sigma_{\xi}$, which regulates the overall level of randomness in the system. 

Combining all these ingredients, the updated angle $\theta^{i}(t)$ is now slightly different from Eq.~\eqref{eq:12}:
\begin{align}
  \label{eq:12a}
\theta^{i}(t) = \phi^{i}(t) \left[1+\xi(t)\left\{1-e^{i}\right\}\right] + \omega^{ij}(t) \Delta \phi^{ij}(t) 
\end{align}
where $\omega(t)$ follows from Eq.~\eqref{eq:16} and 
 $\mathcal{G}\left[\mathbf{o}^{i}(t),\mathbf{o}^{j}(t)\right]$ is still given by Eq.~\eqref{eq:11}.

\subsection{Results of agent-based simulations}
\label{sec:results-agent-based-3}

\begin{figure}[htbp]
  \includegraphics[width=.29\linewidth]{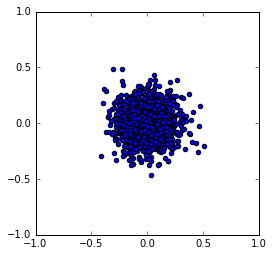}(a)
  \hfill
  \includegraphics[width=.29\linewidth]{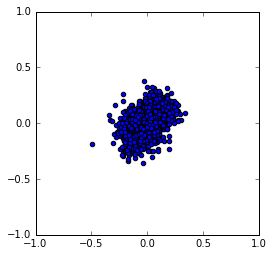}(b)
  \hfill
\includegraphics[width=.29\linewidth]{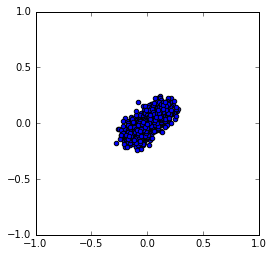}(c)

\includegraphics[width=.29\linewidth]{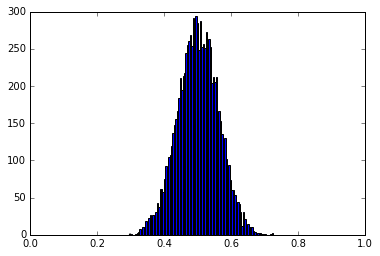}
\hfill
\includegraphics[width=.29\linewidth]{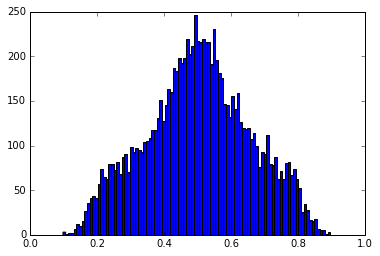}
\hfill
\includegraphics[width=.29\linewidth]{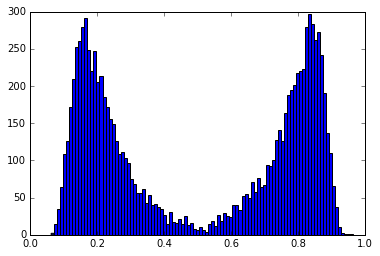}$\quad$
\caption{Opinions of $N=2500$ agents in a multi-dimensional opinion space ($M$=28) at different time steps:
  (a) $t$=0, (b) $t$=380.000, (c) $t$=500.000.
(Top row) The projection of the opinions on the space of the two principal components $\mathbf{c}_{1}$, $\mathbf{c}_{2}$ is shown.
  (Bottom row) Distribution of the corresponding pairwise directional similarity $P[D^{ij}(t)]$, Eq.~\eqref{eq:10}.
Further parameters: $\mu_e = 0.6$, $\sigma_e = 0.5$ for the affective involvement,   $\sigma_{\xi} = 0.2$ for the noise. 
}\label{fig:model3_run}
\end{figure}

In Figure \ref{fig:model3_run} we present the results for the multi-dimensional opinion space with $M=28$, using the same initial setup as before.
We first highlight that our opinion dynamics model is indeed able to produce an outcome with a \emph{bimodal} pairwise directional similarity distribution.
This is achieved despite the fact that the initial distribution, because of the high dimensionality, is unimodal and quite narrow, as shown in  Figure \ref{fig:model3_run}(a).
Secondly, we note that this outcome is \emph{robust} because it is achieved even in those cases where a broader distribution $P[e]$ and an increased noise level are considered.
Different from the case discussed above, this does not lead to consensus in the end.
Instead, we clearly observe the emergence of polarized opinions along the ``ideological dimension''.

The emergence of this dimension out of a high-dimensional opinion space with $M=28$ is indeed remarkable.
It becomes visible in Figure \ref{fig:model3_run} because, different Figure~\ref{fig:aligned_polarization} where only two dimensions exist, here we have chosen to project the positions of agents in the opinion space on the first two principal components, i.e. the dominating dimensions which are explained in more detail in the following Section. 
Because of this projection, the distribution of initial opinions in Figure \ref{fig:model3_run}(a) looks a bit more clustered than in the corresponding Figure~\ref{fig:2d}(a) for the two-dimensional case.

\section{Alignment of opinions and affective involvement}
\label{sec:alignm-princ-comp}

\subsection{Global alignment}
\label{sec:prop-multi-agent}

In this Section, we further explore the emergence of a dominant opinion dimension in a $M$-dimensional opinion space.
One way to extract this dominant dimension from the simulated data is the \emph{principal component analysis} (PCA).
By means of an orthogonal transformation of the original opinions, $\mathbf{o}^{i}(t)=o^{i}_{1}(t)\mathbf{s}_{1}+o^{i}_{2}(t)\mathbf{s}_{2}+\ldots+o^{i}_{M}(t)\mathbf{s}_{M}$, in the opinion space $\mathbf{s}_{1},\ldots,\mathbf{s}_{M}$, it identifies the principal components $\mathbf{c}_{1},\ldots,\mathbf{c}_{M}$, i.e. the axes of a transformed opinion space, such that $\mathbf{c}_{1}$ shows (``explains'') the largest variance in the data, $\mathbf{c}_{2}$ the second largest, etc.
In most cases, already the first few principal components $m=1,2,3$ are sufficient to explain the variance observed.
Compared to the $M$ dimensions, this means a  dimensionality reduction, which necessarily implies a loss of information.
But PCA is a way to \emph{minimize} this loss of information, provided that certain assumptions, for instance about linear correlations between the opinions, hold.

In the following, we only use the first component, $\mathbf{c}_{1}$, based on the insight that the main ideological, or ``left-right'' dimension was found to be the most important one explaining real opinion distributions on policy issues (see also Section~\ref{sec:dimens-reduct-opin}).
In order to calculate how much of the variance is explained by $\mathbf{c}_{1}$ we have to follow the standard procedure of PCA, just summarized here: (a) center the data, i.e. $o^{i}_{1}-\mean{o}_{1}$ etc., (b) compute the covariance matrix, (c) calculate the eigenvalues and corresponding eigenvectors, (d) normalize the eigenvectors to unit vectors, and (e) transform the covariance matrix into a diagonal matrix.
The diagonal elements of this matrix then give us the variance explained by the corresponding axes.
That means, the largest eigenvalue $\lambda_{1}$ refers to the variance explained by the first principal component $\mathbf{c}_{1}$.
In the following we define this as our measures of \emph{global alignment}, $A$:
\begin{align}
  \label{eq:18}
A=\mathrm{Var}(\mathbf{c}_{1})=\lambda_{1}
\end{align}
If  $A=1$, all individual opinion vectors lie on the dominant ideological dimension. 
The lowest value global alignment can attain is $A=1/M$, meaning that there is no global alignment whatsoever.
In the following, we use the term 'global alignment', to distinguish it from individual alignment, which we are going to define further below.
Here, just note that $A$ is \emph{not} defined as an average over individual alignments.

\begin{figure}[htbp]
        \includegraphics[width=0.45\textwidth]{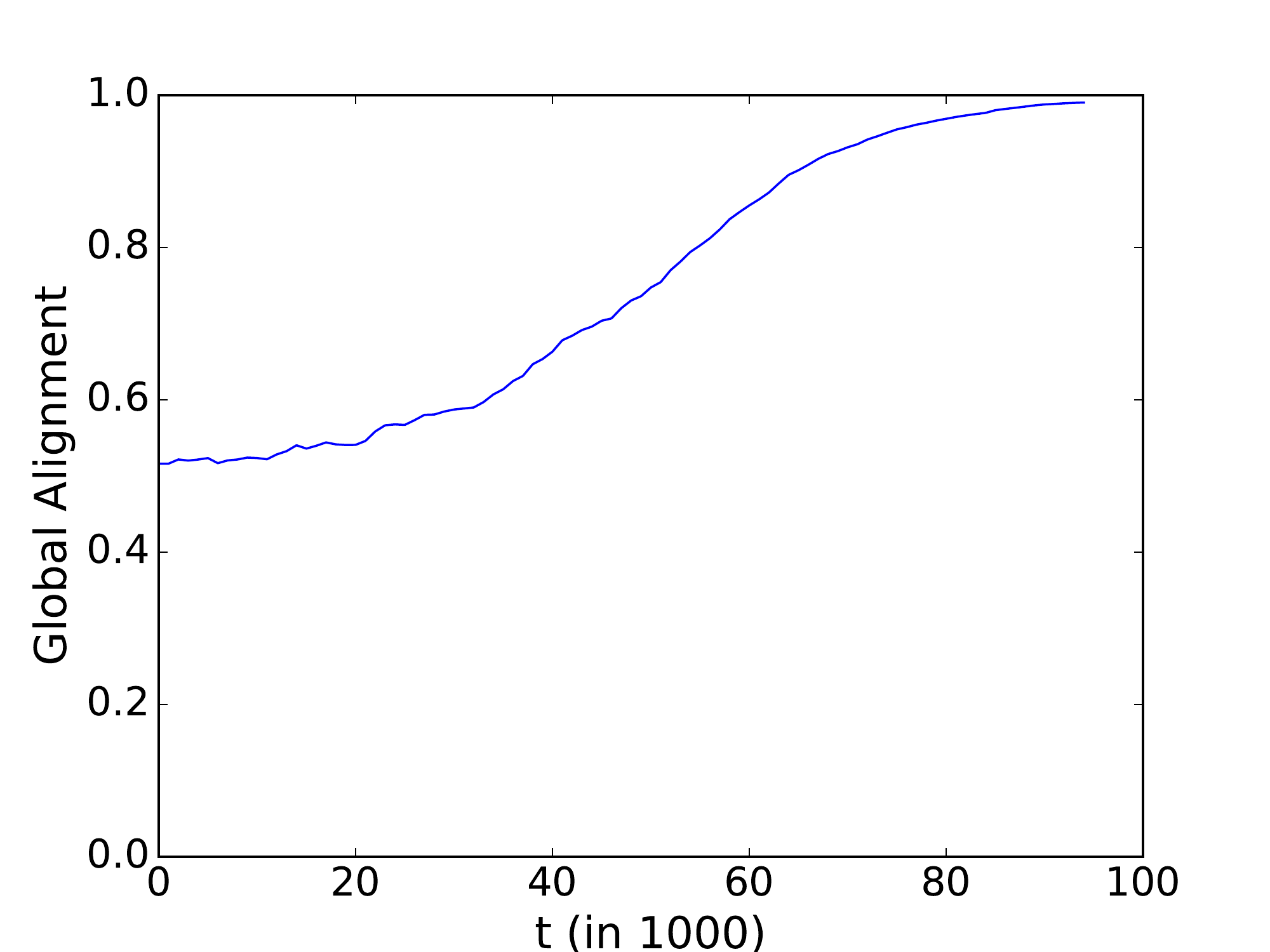}(a)
\hfill
\includegraphics[width=0.45\textwidth]{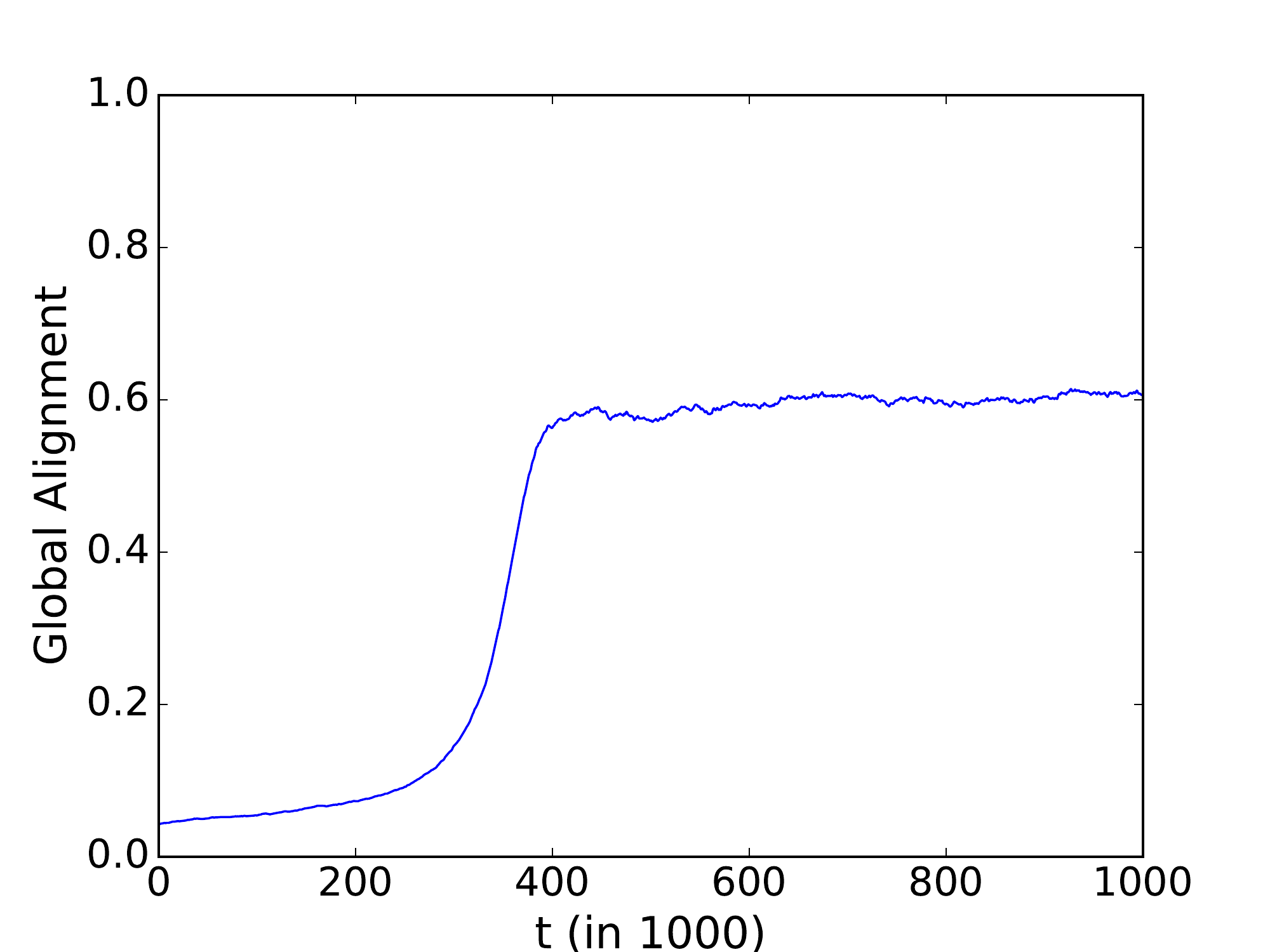}(b)
        \caption{Global alignment $A(t)$ over time measured in simulation steps. (a) Model with opinion alignment, Eqs.~\eqref{eq:10}, \eqref{eq:11},~\eqref{eq:13} and no noise, for $M=2$, also shown in Figure~\ref{fig:aligned_polarization}. (b) Model with opinion alignment and repulsion, Eqs.~\eqref{eq:11}, \eqref{eq:15}, \eqref{eq:12a}, for $M=28$, also shown in Figure~\ref{fig:model3_run}. 
}
    \label{fig:model_2_3-pca}
\end{figure}
Figure~\ref{fig:model_2_3-pca} illustrates how the global alignment $A$ evolves over time for the two agent-based simulations shown in Figures~\ref{fig:aligned_polarization}, \ref{fig:model3_run}.
Figure~\ref{fig:model_2_3-pca}(a) shows the almost perfect global alignment along the dominating dimension, for $M=2$.  
In Figure~\ref{fig:model_2_3-pca}(b), we clearly see that
due to the noise the model never comes completely to rest, and never fully aligns to the first component of the PCA, because of the high number of dimensions, $M=28$. 
However, after 400,000 iterations, the global alignment stabilizes around a relatively high value of 0.6.

\subsection{Impact of affective involvement}
\label{sec:impact-affect-involv}

We now investigate how the global alignment $A$ depends on the second variable that characterizes each agent, the affective involvement $e^{i} \in [0,1]$.
We focus on the opinion dynamics with alignment and repulsion and concentrate on the high-dimensional opinion space, $M=28$.
A discussion of how the affective involvement impacts the model with only aligment was already given in Section~\ref{sec:results-agent-based-3}.

We recall that the value of $e^{i}$ is constant over time, but in general drawn from a truncated normal distribution $\mathcal{N}(\mu_{e},\sigma_{e})$.
That means, varying $\mu_{e}$ between 0 and 1 allows us to increase the expected value for the affective involvement, which  decreases the ability to randomly change the opinion as it directly impacts $\mathcal{Z}[\cdot]$, Eqn.~\eqref{eq:14aa}. 
Varying $\sigma_{e}$ between 0 and 1, on the other hand, allows to make agents more \emph{heterogeneous} regarding this ability.
Thus, high values of $\mu_{e}$ combined with low values of $\sigma_{e}$ would refer to a \emph{deterministic} limit, while low values of $\mu_{e}$ combined with low values of $\sigma_{e}$ would refer to a \emph{random} limit, where all agents are impacted by the noise in the same (large) manner.

In our agent-based computer simulations, we vary both $\mu_e$ and $\sigma_e$ in steps of 0.1.
For each combination $(\mu_e,\sigma_e)$ we run 10 simulations for for 1,000,000 time steps, to ensure that a quasistationary global alignment is reached (see also Figure~\ref{fig:model_2_3-pca}b).
Due to the noise, the simulations never reach a completely stationary state.
To determine the values of $\xi^{i}$, we sample from the truncated normal distribution $\mathcal{N}(\mu_{\xi},\sigma_{\xi})$ with $\mu_{\xi}=0$ and $\sigma_{\xi}=0.2$.

\begin{figure}[htbp]
        \includegraphics[width=0.45\textwidth]{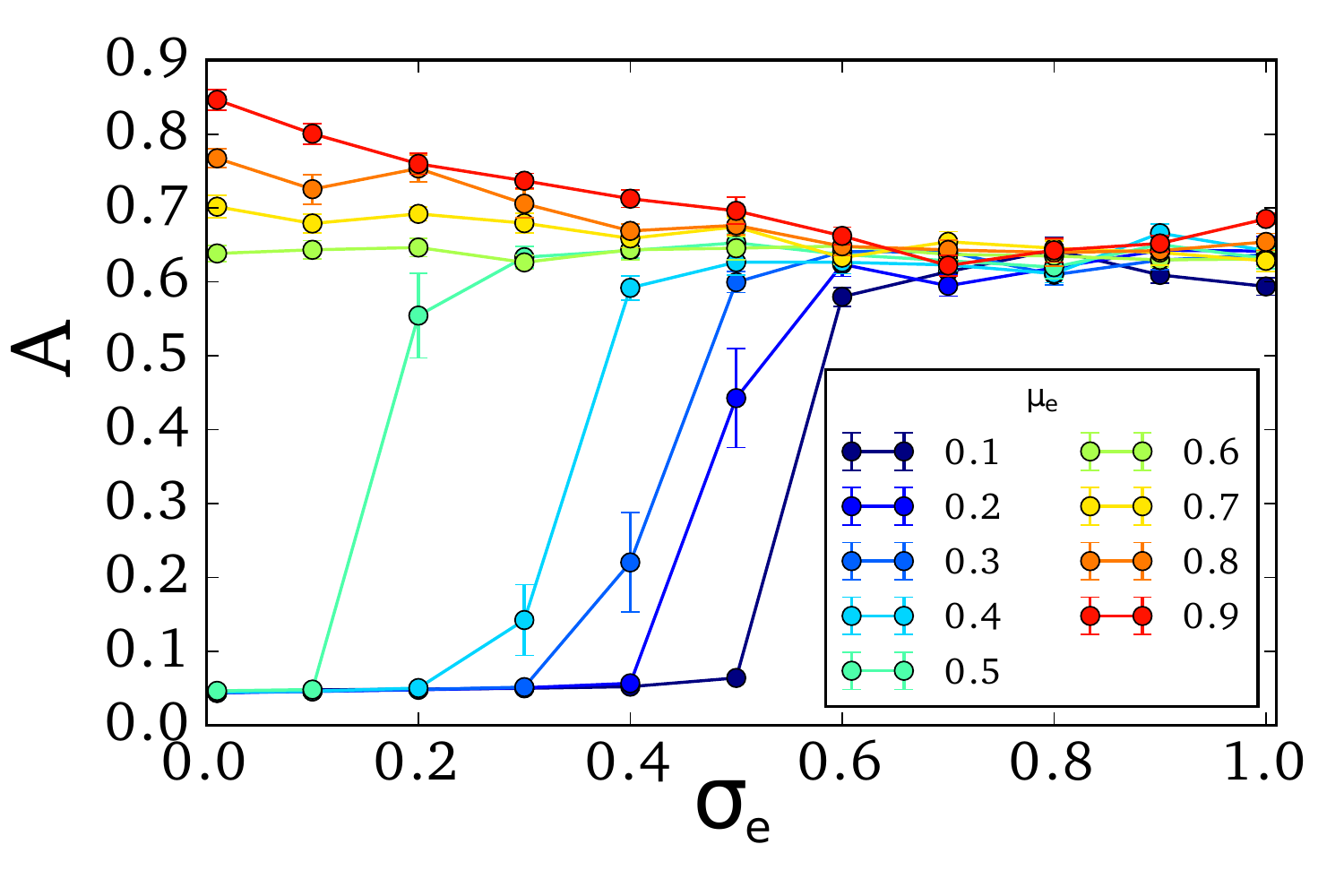}(a)
\hfill
        \includegraphics[width=0.45\textwidth]{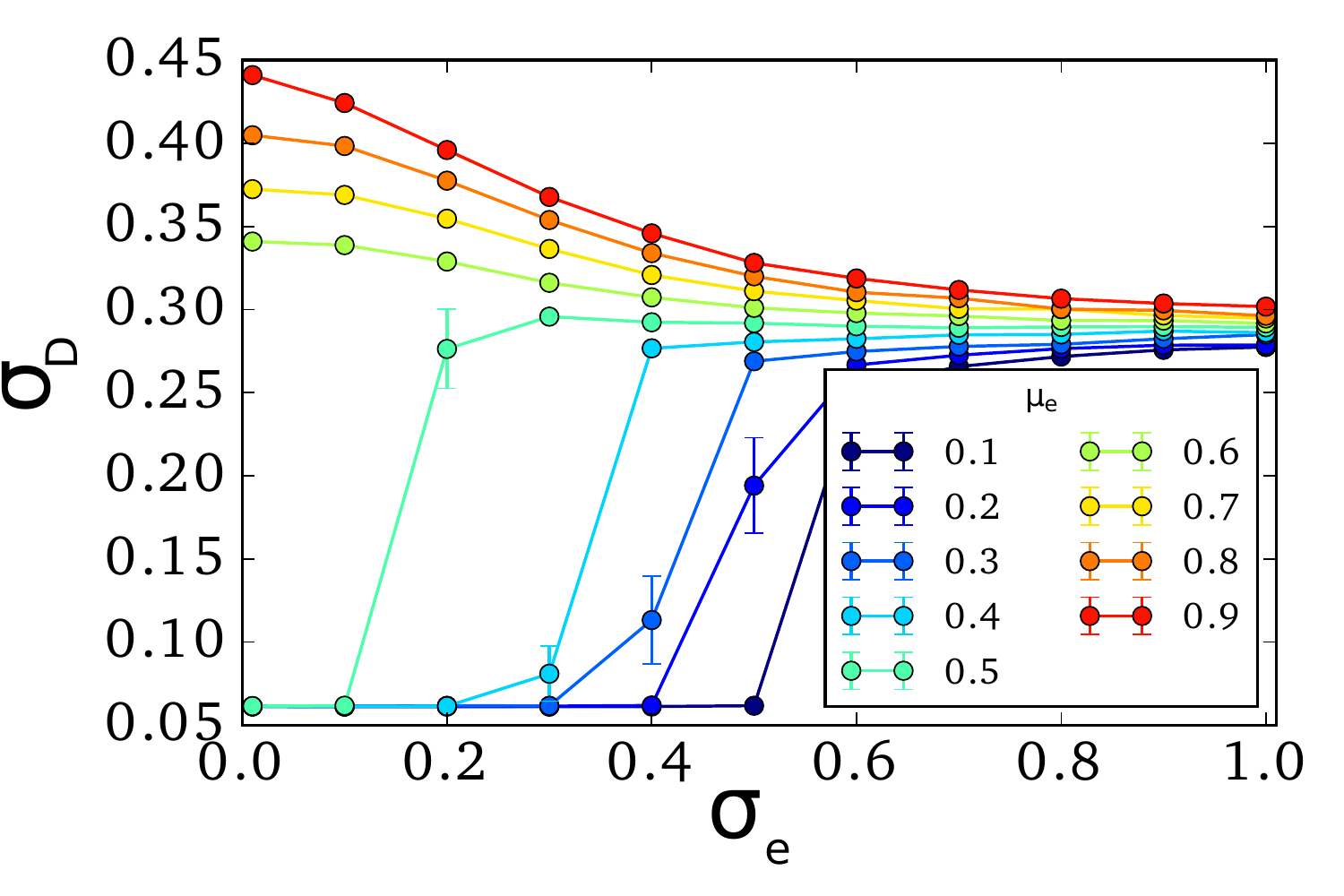}(b)
        \caption{(a) Global Alignment $A$ for varying distributions $\mathcal{N}(\mu_{e},\sigma_{e})$ to describe the affective involvement of agents. Different lines refer to different values of $\mu_{e}$, the $x$-axis to different values of $\sigma_{e}$.
          (b) Standard deviation $\sigma_{D}$ of the distribution $P[D^{ij}]$ of pairwise directional similarities. Error bars indicate the standard errors with respect to the mean.}
  \label{fig:global}
\end{figure}

The results are shown in Figure~\ref{fig:global}(a).
We clearly see that for large $\mu_{e}$, i.e. a low overall level of randomness in the dynamics, the global alignment $A$ is always high.
In the deterministic limit it reaches a level of 85\%, and even for a large heterogeneity in the agents' affective involvement it is still above 60\%.
This contrasts with the random limit of small $\mu_{e}$, where the global alignment drops to zero if $\sigma_{e}$ is below 0.5.
If it is above 0.5, i.e. if by chance there are still sufficiently many agents with a larger affective involvement, this again allows a global alignment of opinions.
Hence, we confirm again that our opinion dynamics with alignment and repulsion is very robust against variations in the agent's parameters.
On the other hand, we find that the transition between aligned and non-aligned global states is rather steep.
That means, there is a critical level of randomness that can destroy the global alignment of opinions, as it should be rightly expected. 

The second variable to characterize the alignment of agents' opinions to the dominant dimension is the \emph{pairwise directional alignment} $D^{ij}(t)=1-\Delta \phi^{ij}(t)/\pi$.
We have shown in Figures~\ref{fig:aligned_polarization}, \ref{fig:model3_run} that the histogram of these values over time approaches a \emph{bimodal distribution}.
This indicates that the system develops a  state of \emph{polarization} where agents form clusters of opinions along opposite directions of the dominant dimension.
That means, there is a \emph{coexistence} of opposite opinions in the long term, whereas a \emph{unimodal} distribution would refer to \emph{coexistence}.
These two outcomes can be characterized by the standard deviation $\sigma_{D}$ of the distribution $P[D]$.
While the mean of this distribution is in both cases $\mu_{D}\approx 0.5$, consensus would refer to \emph{small} values of $\sigma_{D}$, while polarization refers to \emph{large} values of $\sigma_{D}$.
We note that, because the values for $D^{ij}$ are bound between 0 and 1, a large value of $\sigma_{D}$ means 0.5. 

We have also investigated how the distribution $P[D]$ depends on the distribution of the affective involvement $\mathcal{N}(\mu_{e},\sigma_{e})$. 
The results are shown in Figure~\ref{fig:global}(b). 
For the deterministic limit of large $\mu_{e}$ and small $\sigma_{e}$, we find high values for $\sigma_{D}$, i.e. a clear polarization.
This even holds if the heterogeneity of the agents' affective involvement $e^{i}$ is increased.
For the random limit of small $\mu_{e}$ and small $\sigma_{e}$, on the other hand, we see that the global polarization is destroyed by the noise, and instead \emph{consensus} is obtained.
This is in line also with our previous discussions that an increased noise level fosters consensus.
Again, we note that the transition between consensus and polarization is a rather steep, i.e. there exists a critical level of randomness.

\subsection{Individual alignment}
\label{sec:individual-alignment}

We also investigate how the affective involvement $e^{i}$ of individual agents impact their \emph{individual} ability to align to the dominating opinion dimension.
For this alignment we can define an angle $\psi^{i}\left[\mathbf{o}^{i},\mathbf{c}_{1}\right]$ between the individual opinion vector $\mathbf{o}^{i}$ and the main PCA component $\mathbf{c}_{1}$. 
Agent $i$ is perfectly aligned to the main ideological dimension if $\psi^{i} = 0$, i.e. the opinion vector points into the direction of the PCA component $\mathbf{c}_{1}$, or if $\psi^{i} = \pi$, i.e. the opinion vector points into the direction \emph{opposite} to $\mathbf{c}_{1}$.
The latter case indicates individual opposition and forms the basis for polarization.
But still, this opposition can be expressed in terms of the ideological dimension.
Thus, we define the \emph{individual alignment} $a^{i}(t)$ of agent $i$ to the main ideological dimension as:
\begin{align}
  \label{eq:17}
a^{i}(t) = \abs{\frac{2\psi^{i}(t)}{\pi} - 1}  
\end{align}
$a^{i}=0$ if the opinion vector $\mathbf{o}^{i}$ is orthogonal to the dominant ideological dimension $\mathbf{c}_{1}$, $a^{i}=1$ if it is either $0$ or $\pi$, pointing in either direction from the origin.

\begin{figure}[htbp]
  \centering \includegraphics[width=0.45\textwidth]{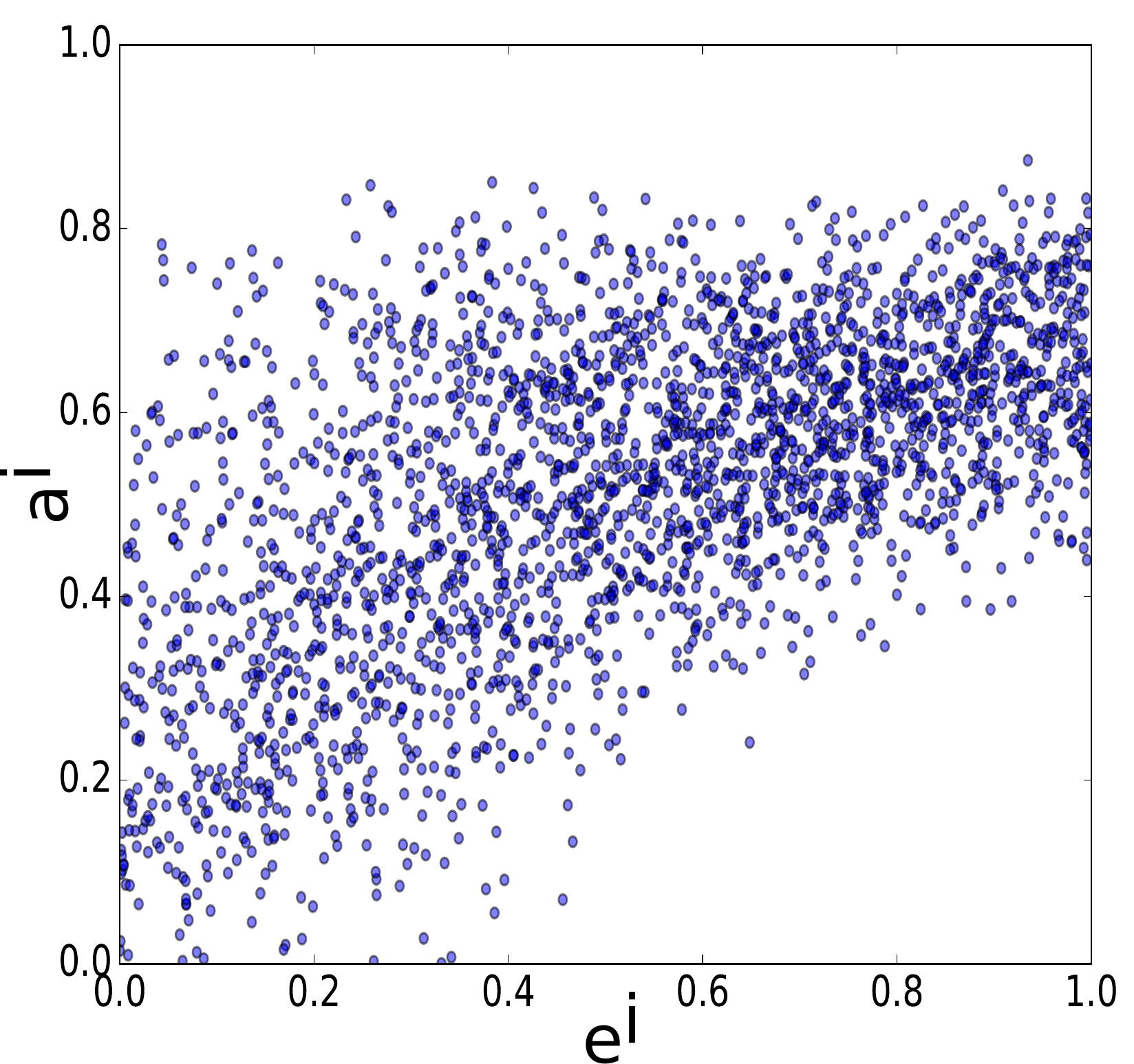}
        \caption{Scatter plot of individual alignment $a_{i}$ vs. emotional involvement $a_{i}$ for the results shown in Figure~\ref{fig:model_2_3-pca}(b) with $M=28$. }
    \label{fig:modeldiagns}
\end{figure}

The scatter plot shown in Figure~\ref{fig:modeldiagns} gives us a first indication of how the two agent variables $e^{i}$ and $a^{i}$ relate.
We see that a higher affective involvement, i.e. a lower level of random opinion changes, indeed correlates with a higher level of individual aligment.
This reflects the dissolving role of noise on alignment, already discussed for the global alignment $A$. 
We note again that $A$ is \emph{not} defined as an average over individual alignments. 

\begin{figure}[htbp]
        \includegraphics[width=0.45\textwidth]{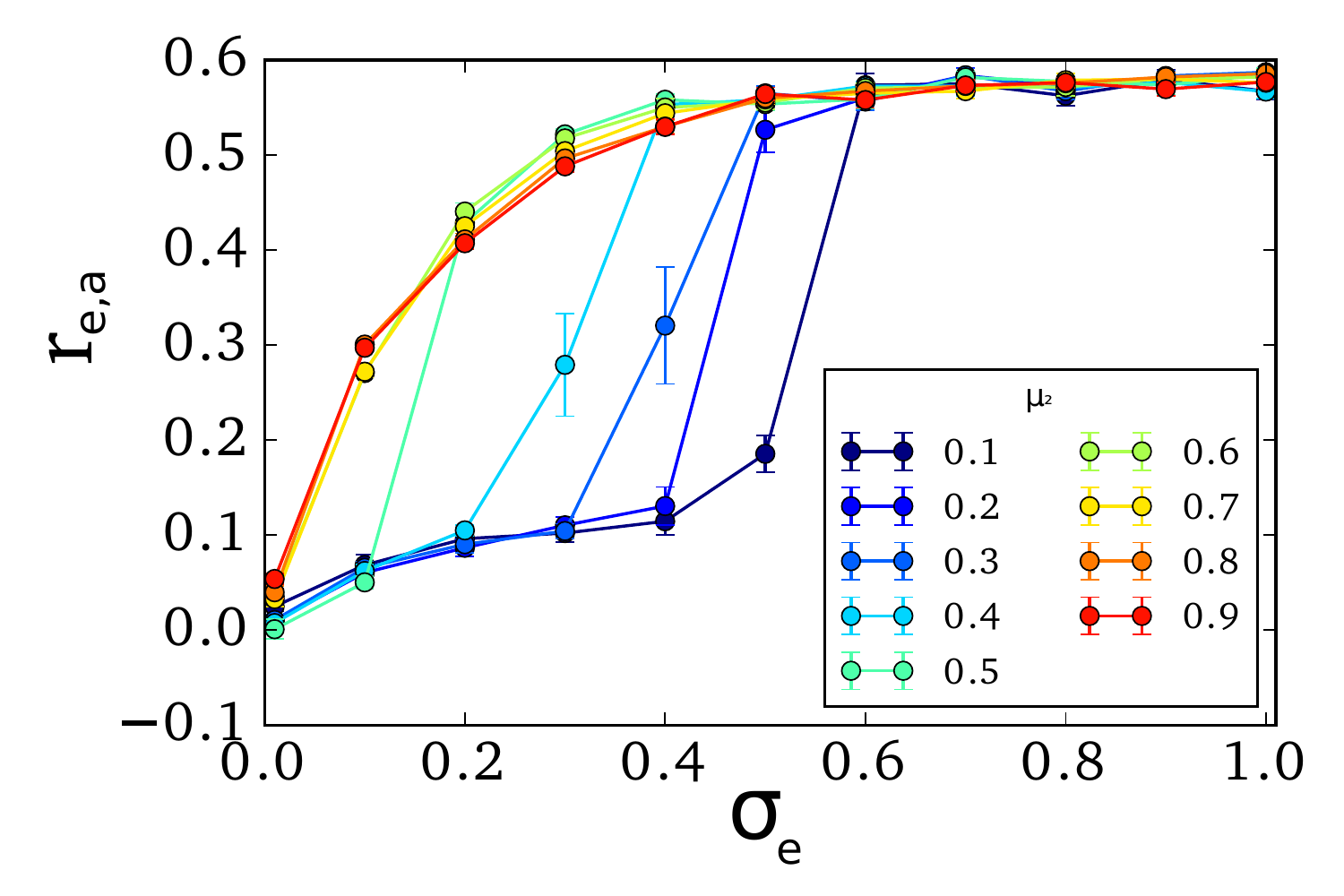}(a)
\hfill
        \includegraphics[width=0.45\textwidth]{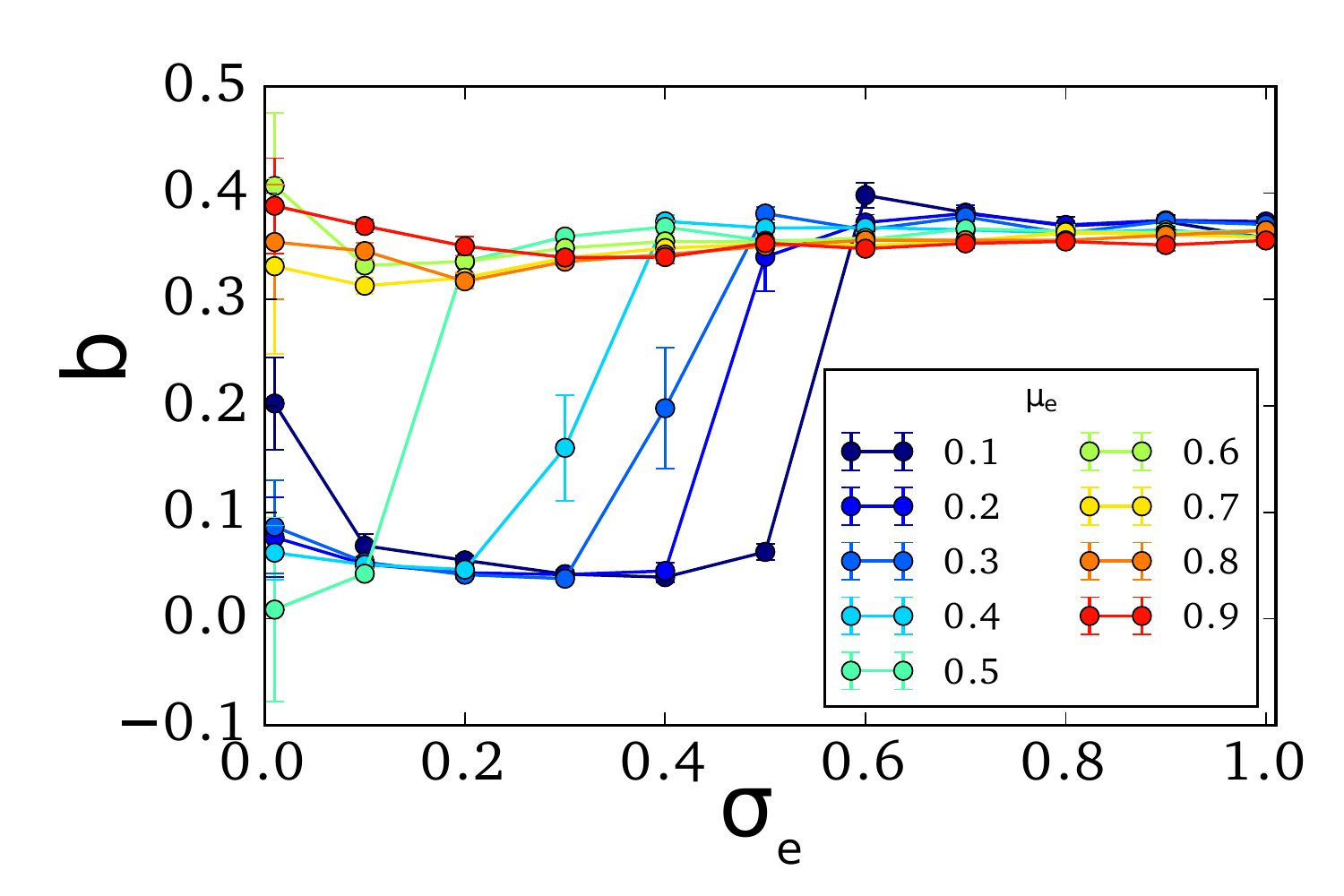}(b)
        \caption{  (a) Correlation $r_{e,a}$, Eq.~\eqref{eq:19} between individual affective involvement $e^{i}$ and  individual alignment $a^{i}$, (b) regression parameter $b$, Eq.~\eqref{eq:20} of $e^{i}$ on $a^{i}$ for varying distributions $\mathcal{N}(\mu_{e},\sigma_{e})$ to describe the affective involvement of agents. Different lines refer to different values of $\mu_{e}$, the $x$-axis to different values of $\sigma_{e}$.
          Error bars indicate the standard errors with respect to the mean.}
    \label{fig:modelruns}
\end{figure}

To study this relation in a more systematic manner, we repeat the simulation procedure used in Section~\ref{sec:impact-affect-involv}. 
We define the Pearson correlation coefficient between $e^{i}$ and $a^{i}$ as:
\begin{align}
  \label{eq:19}
  r_{e,a}=\frac{N\sum_{i}e^{i}a^{i }- \sum_{i}e^{i}\sum_{i}a^{i}}{\sqrt{N\sum_{i}\left(e^{i}\right)^{2}-\left(\sum_{i}e^{i}\right)^{2}}
  \sqrt{N\sum_{i}\left(a^{i}\right)^{2}-\left(\sum_{i}a^{i}\right)^{2}}}
\end{align}
We then vary the distribution of affective involvement $\mathcal{N}(\mu_{e},\sigma_{e})$ from which the $e^{i}$ are sampled, to see how this impacts $r_{e,a}$.
The results are shown in Figure \ref{fig:modelruns}(a).
We find again that in the random limit this correlation breaks down.
For the deterministic limit of large $\mu_{e}$, we see that the correlations also decrease with $\sigma_{e}$.
This is quite obvious because $\sigma_{e}\to 0$ means that all agents have the same affective involvement $e^{i}\to e$.
Their individual alignment $a^{i}$ may still vary, but its correlation with a constant $e$ is zero.

Eventually, we can also analyze the scatter plot of Figure~\ref{fig:modeldiagns} by means of a linear regression model :
\begin{align}
  \label{eq:20}
  a^{i}=c+ b\,e^{i} +\eps^{i}
\end{align}
where $\epsilon^{i}$ is the error term (residual) and $c$ is the intercept. 
The regression parameter $b$ varies with the parameters of the distribution $\mathcal{N}(\mu_{e},\sigma_{e})$ as shown in Figure~\ref{fig:modelruns}(b).
Again, we see that in the random limit of small $\mu_{e}$ $b$ is close to zero, i.e. correlations are low, while in the deterministic limit its value is reasonably large.
The phase transition at a critical $\sigma_{e}$ is also clearly visible.

\section{Conclusion}
\label{sec:abm_conclusion}

\paragraph{Agent-based modeling. \ }

In this paper, we apply agent-based modeling as one possible methodology to understand an empirically observed macro phenomenon, in our case the alignment of individual opinions.
Agent-based modeling requires us to provide reasonable \emph{micro mechanisms} of how agents influence another in their opinions.
We base the proposed mechanisms in established \emph{psychological theories}, notably cognitive dissonance theory and structural balance theory.

Our model then allows to test how different assumptions about such mechanisms impact the macroscopic dynamics.
That means, we do \emph{not} follow a data-driven modeling approach, which tries to reproduce a specific real-world outcome by estimating interaction parameters from observations \citep{Schweitzer2018datadriven}. 
Instead, we aim at  a ``generative explanation'' \citep{epstein2006generative}, a thought experiment to find out which mechanisms are necessary and sufficient to generate a stylized version of empirical reality -- and, perhaps even more important, which mechanisms are \emph{not} sufficient. 
By adding or removing mechanisms and tuning model parameters we can improve our model step by step, until we finally attain a model that is able to reproduce the desired macro-phenomenon.
This of course does not proves the validity of the model assumptions, but it is a clear indicator which modeling hypotheses are compatible with a given macroscopic outcome.

\paragraph{Obtaining global issue alignment and polarization. \ }

Our model shall be able to reproduce two features of opinion dynamics observed in the political domain:
(i) The emergence of global issue alignment: Individual opinions on different policy issues are correlated such that a dominant ``left-right'' ideological dimension can explain most of them. (ii) A polarization of opinions on this ideological dimension:
While individual opinion vectors are aligned, they point to opposite directions from the origin.

Global issue alignment already assumes an underlying \emph{multi-dimensional opinion space}, which is neglected in many opinion dynamics models.
It means that instead of a scalar value, opinions are characterized by \emph{vectors} in this $M$-dimensional space. 
The existence of global alignment was empirically demonstrated, it was also theoretically discussed in political science with respect to
voting and coalition formation \citep{baldassarri2008partisans,merrill1999unified,Indridason2011}.
However, there is still a lack of models that are able to generate this phenomenon.

We fill this research gap by investigating the conditions under which global issue alignment is obtained.
Specifically, in our model we vary (a) what information about the opinions of others agents take into account, and (b) how they respond to this information.
We have shown that the so called \emph{proximity voting}, which is equivalent to using the Euclidean distance to evaluate the similarity of opinions, fails to generate a global issue alignment.
\emph{Directional voting}, however, in which agents measure similarities of opinions dependent on the ``right'' and ``wrong'' side, has the potential to generate global alignment, at least in low-dimensional opinion spaces.
If we combine directional voting with a repulsive force between far-distant opinions, we find that global issue alignment also emerges in high-dimensional opinion spaces and is very robust against parameter changes in the model.

The repulsive force is not just postulated to improve the model, it is motivated by the mentioned psychological mechanisms, structural balance  and minimization of cognitive dissonances,  if agents have different opinions on a given issue.
This could either lead to an attractive force, i.e. opinions of agents become more similar to minimize the dissonance, or to a repulsive force, i.e. opinions of agents become more different.
Hence, we propose a reasonable micro mechanism.
Even more, we also demonstrate how these assumptions can be formalized in an opinion dynamics model, by providing a formal model in polar coordinates.

\paragraph{Affective involvement. \ }

As an asset, our model includes an emotional component in the opinion dynamics. 
This extension is rooted in arguments from political science theory that global issue alignment is driven by `\emph{passion}', i.e., affective involvement in politics \citep{poole2005spatial}.
We implement the emotional component in our agent-based model by means of a heterogeneous parameter, $e^{i}$, drawn from a distribution $\mathcal{N}(\mu_{e},\sigma_{e})$ with given mean and variance.
The higher the level of affective involvement, the more resistant are agents to change their opinions.
We discuss two different ways of implementing this relation, (a) by defining a threshold for directional similarity, (b) by impacting the level of random opinion changes.

In our paper, we systematically study the influence of affective involvement on the global issue aligment and on the individual alignment.
We find that  two different types of outcome can be observed:
If the level of affective involvement, expressed by $\mu_{e}$, is low and the heterogeneity across agents, expressed by $\sigma_{e}$ is also low, we end up in  regime with  little global alignment, low opinion polarization, and little  individual alignment.
In this \emph{disorganized} state, no dominant ideological dimension emerges to which agents align their opinions.

If on the other hand, the heterogeneity of agents' affective involvement is high, which means that (for both low and high $\mu_{e}$) there is a sufficiently large number of agents with a \emph{high level of affective involment}, we always find outcomes with high global issue alignment, high opinion polarization and high individual alignment.
This is a highly \emph{organized} state in opinion space, and we have pointed out that there is a rather sharp transition between the disorganized and the organized states dependent on the parameters of the affective involvement.  
So, we can conclude that affective involvement, the way it is considered in our model, fosters the global issue alignment, as argued also by political scientists.
Even more, global aligment can be only observed beyond a critical level of affective involvement.

\small
\setlength{\bibsep}{1pt}

\bibliography{add,ref2}

\appendix
\section*{Appendix}
\label{sec:appendix}

\section{Simulation of the two-dimensional bounded confidence model} 
\label{sec:simul-two-dimens}

\begin{figure}[htbp]
  \begin{center}
   \includegraphics[width=.29\linewidth]{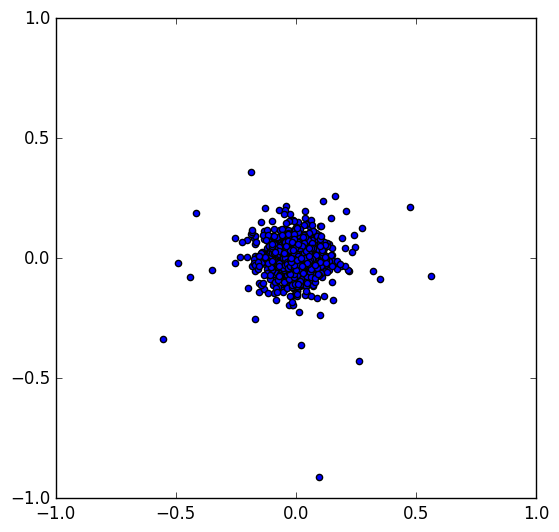}(a)
$\quad$
\includegraphics[width=.29\linewidth]{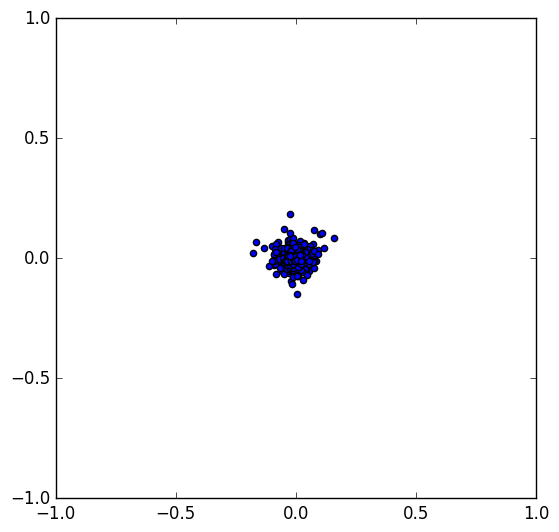}(b)

\includegraphics[width=.29\linewidth]{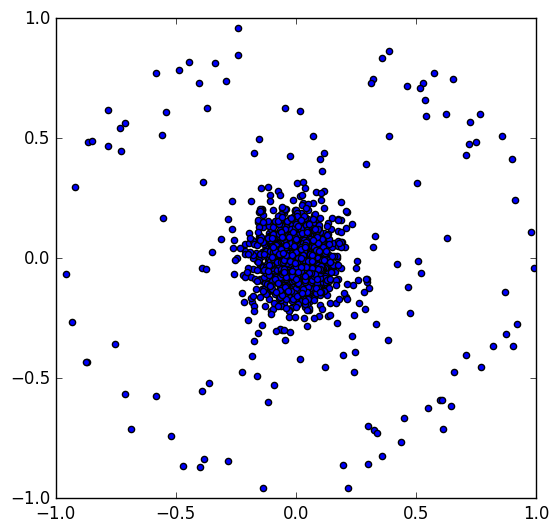}(c)
$\quad$
\includegraphics[width=.29\linewidth]{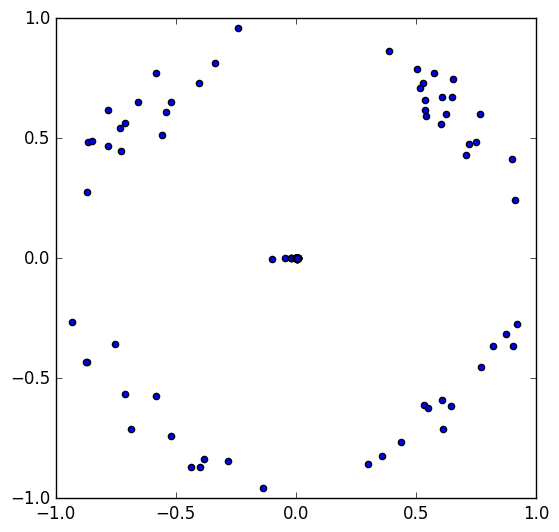}(d)
\end{center}
\caption{Snapshots of the positions of agents in the two-dimensional opinion space:  (a,b) $\epsilon=0.5$, (c,d) $\epsilon=0.25$, 
(a) $t$=30.000, (b) $t$=40.000, (c) $t$=30.000, (d) $t$=90.000. 
}
  \label{fig:evolution}
\end{figure}

Here we present snapshots of the dynamics, for which the initial state is shown in Figure~\ref{fig:2d-bounded}(a) and the final states in Figures\ref{fig:2d-bounded} (b)and (c). 
The model parameters are chosen as follows:  $\mu_{o}=0$ and $\sigma_{o} = 0.8$ for the initial opinions, 
$\mu_e =0.5$ and $\sigma_e = 0$ for the emotional involvement, i.e., all agents have the same $e^{i}\equiv e=$0.5.
For the confidence interval, two different values are chosen: $\epsilon=0.5$ in Figures \ref{fig:evolution}(a,b),  $\epsilon=0.25$ in Figures \ref{fig:evolution}(c,d).

\begin{figure}[htbp]
  \includegraphics[width=.29\linewidth]{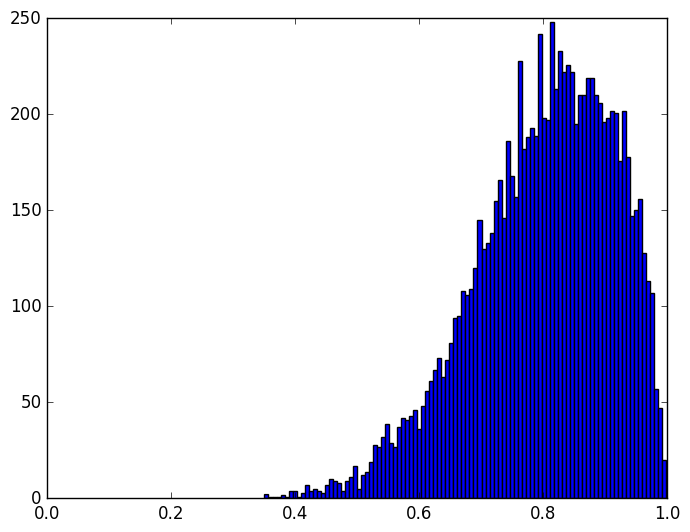}(a)
  \hfill
  \includegraphics[width=.29\linewidth]{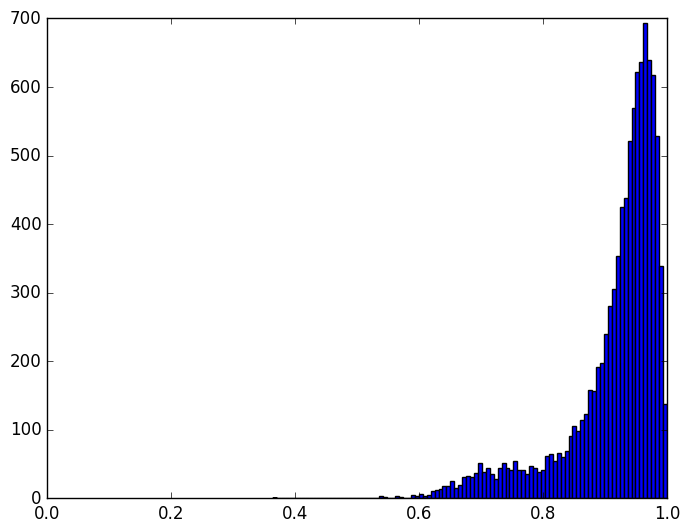}(b)
  \hfill
\includegraphics[width=.29\linewidth]{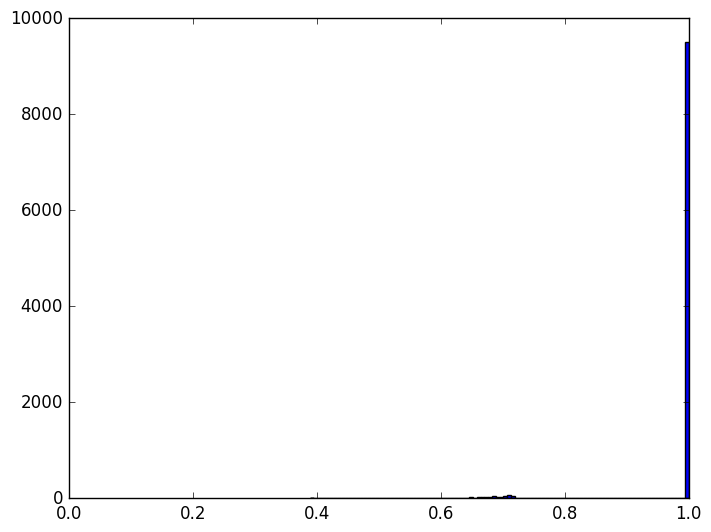}(c)
  \caption{Distribution of the pairwise Euclidean similarity $S^{ij}(t)$, Eq.~\eqref{equ:similarity_1}, at different time steps: (a) $t$=0, for the distribution in opinion space see Figure~\ref{fig:2d}(a),  (b) $t$=30.000, (c) $t$=210.000. Parameters see Figure~\ref{fig:evolution}, $\epsilon=0.25$. }
  \label{fig:similarity-bounded}
\end{figure}

To further quantify the dynamics of the agent-based model, we analyze the evolution of the 
\emph{pairwise Euclidean similarity}:
\begin{equation}
{S}^{ij}(t) =  1 - \frac{d^{ij}(t)}{\sqrt{8}}
\label{equ:similarity_1}
\end{equation}
$S^{ij}(t)$ is a linear transformation of the pairwise Eucledian distance and is shown in Figure~\ref{fig:similarity-bounded}.
We see that initially the distribution $P[S]$ is rather broad, but becomes more narrow over time, to converge almost to a delta peak at $S=1$.
That means that the pairwise similarity is maximized for all agents. 
In case of consensus, this happens because \emph{all} agents have reached the same opinion vector, also shown in Figure\ref{fig:2d}(b).
In case of coexistence, the outcome is almost identical because the opinion clusters in the periphery, also shown in Figure\ref{fig:2d}(c), contain only very few agents.
Hence, the very small contribution at about $S=0.7$ in \ref{fig:similarity-bounded}(c) is barely noticeable.
After all, this configuration is stable because \emph{those} agents that still interact, have reached the same opinion vector and belong to the same cluster in the opinion space.

\section{Simulation of the opinion alignment in $M=28$ dimensions}
\label{sec:model-2-consensus}

\begin{figure}[htbp]
\centering
\includegraphics[width=.29\linewidth]{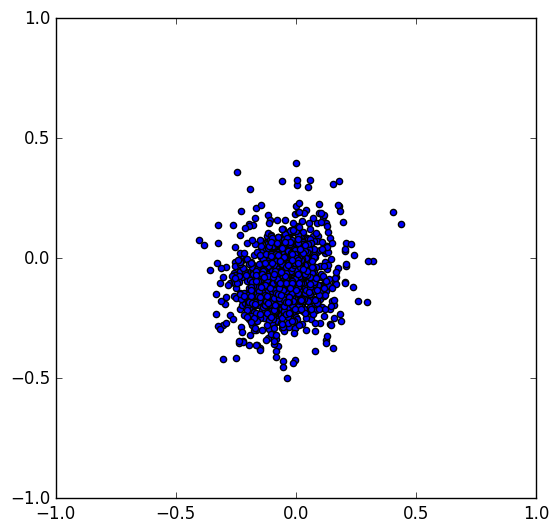}(a)
  \hfill
  \includegraphics[width=.29\linewidth]{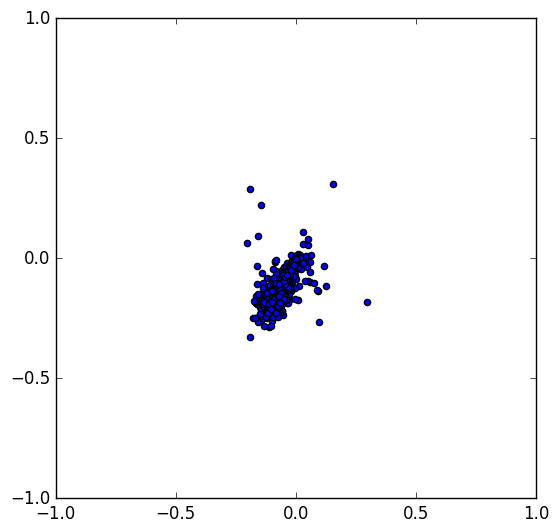}(b)
  \hfill
  \includegraphics[width=.29\linewidth]{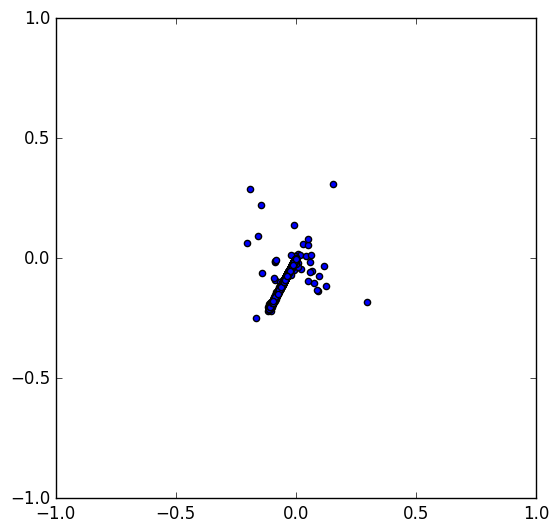}(c)\\
\includegraphics[width=.29\linewidth]{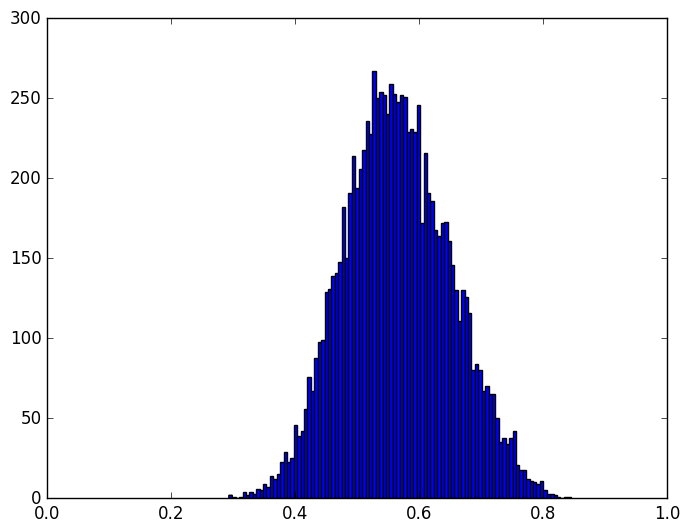} \hfill
\includegraphics[width=.29\linewidth]{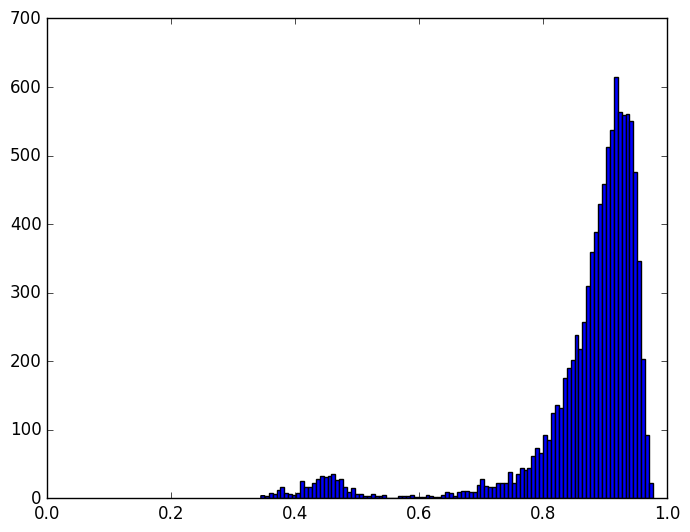} \hfill
\includegraphics[width=.29\linewidth]{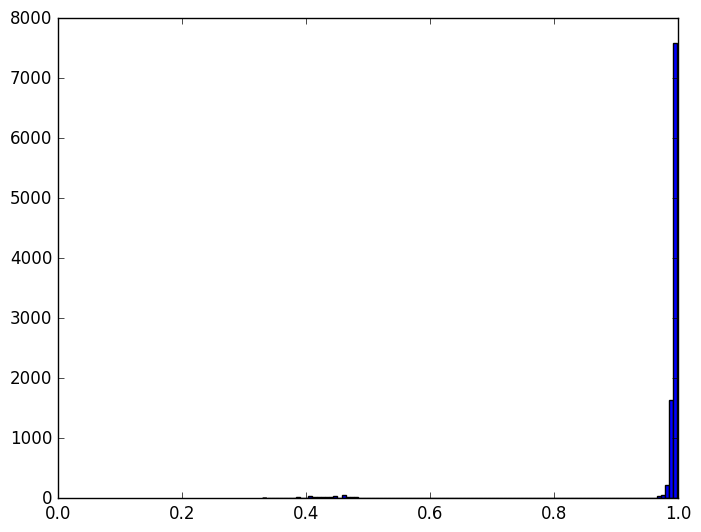}$\quad$
\caption{Opinions of $N=2500$ agents in a multi-dimensional opinion space ($M$=28) at different time steps:
 at different time steps: (a)$t$=50.000, (b) $t$=70.000,  $t$=100,000.
(top row) The projection of the opinions on the space of the two principal components $\mathbf{c}_{1}$, $\mathbf{c}_{2}$ is shown.
(bottom row) Distribution of the pairwise directional similarity, $P[D^{ij}(t)]$, Eq.~\eqref{eq:10} (c)  $P[D^{ij}(0)]$ is shown in Figure~\ref{fig:initial_distribs}(c). Further parameters: $e^{i}\equiv e=0.5$, no noise.
  }\label{fig:28_D}
\end{figure}

In Figure~\ref{fig:28_D} we present results of the multi-dimensional opinion alignment without repulsion.  
The results are discussed in Section~\ref{sec:high-dimens-opin}.

\end{document}